\documentclass[journal,twoside]{asmb}

\begin{document}
\title{Enhancing Security of TAS/MRC Based Mixed RF-UOWC System with Induced Underwater Turbulence Effect}

\author[1]{Md. Ibrahim}
\author[2]{A. S. M. Badrudduza}
\author[3]{Md. Shakhawat Hossen}
\author[4]{M. K. Kundu}
\author[5]{Imran Shafique Ansari}

\affil[1]{Department of Electrical \& Electronic Engineering, Rajshahi University of Engineering \& Technology (RUET), Rajshahi-6204, Bangladesh}
\affil[2,3]{Department of Electronics \& Telecommunication Engineering, RUET}
\affil[4]{Department of Electrical \& Computer Engineering, RUET}
\affil[5]{James Watt School of Engineering, University of Glasgow, Glasgow G12 8QQ, United Kingdom}

\twocolumn[
\begin{@twocolumnfalse}
\maketitle
\begin{abstract}
\section*{Abstract}

Post commercial deployment of fifth-generation (5G) technologies, the consideration of sixth-generation (6G) networks is drawing remarkable attention from research communities. Researchers suggest that similar to 5G, 6G technology must be human-centric where high secrecy together with high data rate will be the key features. These challenges can be easily overcome utilizing PHY security techniques over high-frequency free-space or underwater optical wireless communication (UOWC) technologies. But in long-distance communication, turbulence components drastically affect the optical signals, leading to the invention of the combination of radio-frequency (RF) links with optical links. This work deals with the secrecy performance analysis of a mixed RF-UOWC system where an eavesdropper tries to intercept RF communications. RF and optical links undergo $\eta-\mu$ and mixture exponential generalized Gamma distributions, respectively. To keep pace with the high data rate of the optical technologies, we exploit the antenna selection scheme at the source and maximal ratio combining diversity at the relay and eavesdropper, while the eavesdropper is unaware of the antenna selection scheme. We derive closed-form expressions of average secrecy capacity, secrecy outage probability, and probability of strictly positive secrecy capacity to demonstrate the impacts of the system parameters on the secrecy behavior. Finally, the expressions are corroborated via Monte-Carlo simulations.
\end{abstract}

\begin{IEEEkeywords}
\section*{Keywords} 

Maximal ratio combining, physical layer security, secure outage probability, transmit antenna selection, under water optics.

\end{IEEEkeywords}
\end{@twocolumnfalse}
]

\section{Introduction}
\subsection{Background and Literature Study}

Over the previous decades, the emergence of underwater activities has been perceived greatly due to its utilization in a variety of applications, such as ecological supervision, oil and gas management system, coastal security, and military underwater vehicles \cite{zeng2016survey,kaushal2016underwater}. Since the conventional radio frequency (RF) and acoustic wireless carriers provide a low speed of data rate along with severe communication delays in underwater communication (UWC) networks, newly, the research community has exhibited significant attraction in alternative options \cite{kaushal2016underwater}. In this regard, underwater optical wireless communication (UOWC) system that allows a large data speed (i.e. tens of Gbps) at average transmission distance and is also applicable to recent sixth-generation (6G) communications \cite{dang2020should}, has been revealing as an optimistic technology \cite{gabriel2013monte}. Moreover, evaluation of optical communication scheme beneath the water environment has been recognized with immense acceptance as it has advantages of secure transmission, small energy consumption, as well as lower latency \cite{khalighi2014underwater}.

{\color{black}In recent times, most of the UOWC turbulence models have been derived from free-space optical (FSO) communication models. However, the effects of turbulence on FSO and UOWC channels are quite different. The reason behind this is presence of interference in UOWC systems due to absorption and scattering \cite{zhang2018capacity}.} So, despite the mentioned encouraging aspects of UOWC, the absorption and scattering of optical signals created by underwater components will affect the transmission of light beams \cite{guerra2017effect}. The authors of \cite{zeng2015survey} performed a strong comparison among possible strategies of UWC (e.g. RF, acoustic, and optical signals) where they showed that optical technology delivers maximum data speed but experiences scattering as well as absorption phenomena. The optical signal propagation in light beams is extremely suffered by underwater turbulence (UWT) conditions \cite{jamali2016statistical}. The experimental study in \cite{oubei2017performance} discovered that the larger size of air bubbles block optical light beams greatly that may degrade the performance of UWC. 

With a view to improving the performance of the UOWC system, the multiuser diversity technique was introduced in \cite{pang2019performance} where UWC links are subjected to a log-normal model. Including channel impulse response (CIR) and path loss in an underwater environment, a complete structure of UOWC link was developed in \cite{boluda2020impulse} while authors in \cite{oubei2017simple} performed a detailed experiment over UOWC link to investigate the effect of weak turbulence occurred due to underwater elements. Assuming the existence of angular pointing errors in oceanic water, the closed-form expression (CFE) of bit error rate (BER) was developed in \cite{boluda2020impact}.

Nevertheless, very little research has been conducted to model severe turbulence impairments in water environments. Natural turbulence that may occur due to Coriolis, tidal, and weather systems, were studied in \cite{ata2021ber}. Interestingly, the authors revealed that UOWC quality degrades with water temperature as well as water salinity and improves with the source wavelength. 
In \cite{7829336}, the performance of the multi-hop UOWC system was analyzed while 

Reference \cite{bhowal2018performance} modeled a two-way relay-based UOWC network where the influence of scattering, absorption, and turbulence scenarios on system performance was considered. Zhang $et. al$ implemented blue and green laser diodes for non-orthogonal multiple access (NOMA) based UOWC system to establish a high data speed network \cite{zhang2020towards}. Several encryption methods were reported in \cite{du2021experimental,wang2021performance} over the UOWC network that reveals the feasibility of the proposed network under several water turbidities and shallow environments. A recent investigation developed mixture exponential-generalized Gamma (mEGG) distributions that characterize various UWT conditions (i.e. air bubble levels and temperature gradients) effectively for both salty and fresh water environments \cite{zedini2019unified}.

Due to limitations in UOWC caused by turbulence conditions, a mixed dual-hop communication network through a relay node is a feasible solution to upgrade system performance and expand the communication range \cite{ansari2015performance}. In recent times, wireless communication between airborne and underwater vehicles are utilized in the mixed RF-UOWC technologies to establish a high-speed communication network
\cite{li2020performance,lei2020performance,ramavath2020co,7883900}. In \cite{lei2020performance}, a mixed RF-UOWC framework was considered to investigate the same under different turbulence scenarios where main findings were reported to the effect that lower level of air bubbles and temperature gradients reduces scattering and absorption phenomena, hence enhances the system performance. Some notable research over oceanic water environment like \cite{ramavath2020co} considered hyperbolic tangent log-normal (HTLN) model as UWC link to observe the impact of pointing error in RF-UOWC system. Analytical expressions of average BER were developed by authors of \cite{anees2019performance} to examine the influence of weak to severe UWT conditions in a two-hop network. Notably, a communication system that connects satellite and underwater vehicles with lower latency was investigated in \cite{8812666}.

With the noteworthy growing interest in utilizing optical technology in UWC systems, enhancement of security in the corresponding wireless framework poses a significant threat for next-generation communications. {\color{black}Interestingly, optical link delivers a higher level of security because of its' extremely directional light beams \cite{9353550}. However, investigation of security in RF-optical based mixed system becomes challenging since RF link can be attacked easily by malicious eavesdroppers due to their broadcasting nature. As a result, considering security concerns, physical layer security (PLS) has been examined thoroughly in RF-UOWC and RF-FSO systems. Although RF-FSO mixed system has been studied in \cite{lei2017secrecy,islam2020secrecy,lei2020secure}, it differs from mixed RF-UOWC system due to their variety of turbulence effects in the optical links.} Several existing works like \cite{illi2018secrecy,illi2018dual,illi2019physical} considered mixture exponential-Gamma (mEG) distribution for the underwater links. In \cite{illi2018secrecy}, secrecy performance was evaluated in terms of intercept probability (IP) while authors showed the impact of using multiple antennas at the relay node. Deploying amplify-and-forward (AF) relaying technique, secrecy characteristics including RF system parameters and water turbulence were analyzed in \cite{illi2018dual} where RF link experiences Nakagmi-$m$ fading model. 
Authors in \cite{illi2019physical} showed the advantages of diversity order utilizing selection combining (SC) and maximal-ratio-combining (MRC) at receiver for regenerative mixed RF-UOW network. They also proved that multiple antenna diversity establishes secure communication by increasing data speed in RF links. But mEGG distribution is the most versatile model that also incorporates mEG scenario. As a result, assuming mEGG model for UOWC link, \cite{badrudduza2021security} investigated the PLS and proposed the consideration of air bubble sizes and temperature gradients in the underwater environments being crucially important due to the enhancement of secrecy performance. {\color{black}It should be noted that \cite{lei2020secure} is the only work that investigated PLS of RF-FSO based mixed system with transmit antenna selection (TAS) scheme over the RF link. Interestingly, they claimed that increasing the number of antennas would not greatly improve the secrecy performance.}

\subsection{Motivation and Contributions}
In order to meet the rapid demands for high data rates with extreme secrecy, multiple-input multiple-output (MIMO) channels are drawing widespread concerns of research communities. {\color{black} In the RF-UOWC mixed system, UW link is capable of transmitting data at a high rate of tens of Gbps \cite{li201716} whereas, on the other hand, data rate of RF link is significantly lower than UW links due to bandwidth and power constraints. Despite the fact that TAS / maximal ratio combining (MRC) system provides all key features of a MIMO system accompanied by reduced cost and complexity, \cite{yang2018physical} and \cite{ansari2021outage} are the only studies that incorporate TAS/MRC scheme in mixed RF-UOWC network. However, the authors only evaluated system performance ignoring security aspect of practical system.} Motivated by emerging technologies for wireless communication and in order to reduce this data-rate mismatch, we propose a secure dual-hop RF-UOWC network in presence of a passive eavesdropper exploiting TAS / MRC (i.e. MIMO) diversity at the source to relay link. Since UW link is highly secure, eavesdroppers with multiple antennas can only utilize another RF link for intercepting confidential data. The RF links undergo $\eta-\mu$ fading whereas the UW link experiences mEGG distribution. It may be noted, although mEGG channel is the most generic model for UOWC link that characterizes the turbulence scenarios effectively under all conditions \cite{zedini2019unified,ansari2010composite,7145711}, there are limited research works that have been conducted utilizing mEGG model within the PLS domain. {\color{black} Along underwater turbulence, the pointing error is also a factor that tends to reduce the systems capacity. Harmful effects of pointing error are also considered in the underwater mEGG model. Note that, to the authors' best knowledge, this is the pioneering work that demonstrates the impact of pointing error for evaluating the secrecy performance over TAS/MRC-based RF-UWOC link}. Hence, the main contributions of this work are listed as follows:

\begin{itemize}
    \item At first, we derive the PDF of end-to-end dual-hop signal-to-noise ratio (SNR) after realizing the PDF of the source to relay SNR with TAS / MRC diversity and relay to destination SNR. Since eavesdropping is passive-type, the eavesdropper is unaware of the TAS scheme. To the best of the authors’ knowledge, conventional works only considered fading around RF link for performance evaluation \cite{illi2018secrecy,illi2018dual,illi2019physical,8742387,badrudduza2021security} while how TAS / MRC scheme around RF link of a dual-hop RF-UOWC mixed network can enhance secrecy performance, has not been focused yet.
    
    \item The secrecy performance has been characterized in terms of average secrecy capacity (ASC), secrecy outage probability (SOP), and probability of strictly positive secrecy capacity (SPSC). As per the authors’ best knowledge, these expressions are novel as the $\eta-\mu$-mEGG based mixed RF-UOWC system has not been reported yet in the literature. Additionally, the proposed model unifies the secrecy performance of a wide range of classical fading scenarios since both $\eta-\mu$ fading and mEGG turbulence models are generalized statistical models. So, the expressions of performance metrics are also generalized.
    
    \item We utilize the derived expressions of ASC, SOP, and probability of SPSC to quantify the impact of TAS / MRC diversity, fading, underwater turbulence (for temperature gradient and thermally uniform fresh and salty waters), etc. {\color{black} The effect of pointing error on the system performance is also extensively studied in this work. Moreover, a comparison between the systems with and without eavesdroppers is also presented.} Finally, all the derived expressions are validated via Monte-Carlo (MC) simulations.
\end{itemize}

\subsection{Organization}
The remainder of this paper is structured as follows. Section II describes the proposed system and channel models whereas analytical expressions of three significant performance metrics e.g. ASC, SOP, and probability of SPSC are derived in Section III. In Section IV, illustrative numerical outcomes are presented briefly. Finally, Section V delineates the concluding remarks followed by future scopes of our investigation.

\section{System Model and Problem Formulation}
 \begin{figure*}[!ht]
   \vspace{-5mm}
       \centerline{\includegraphics[width=0.55\textwidth,angle =0]{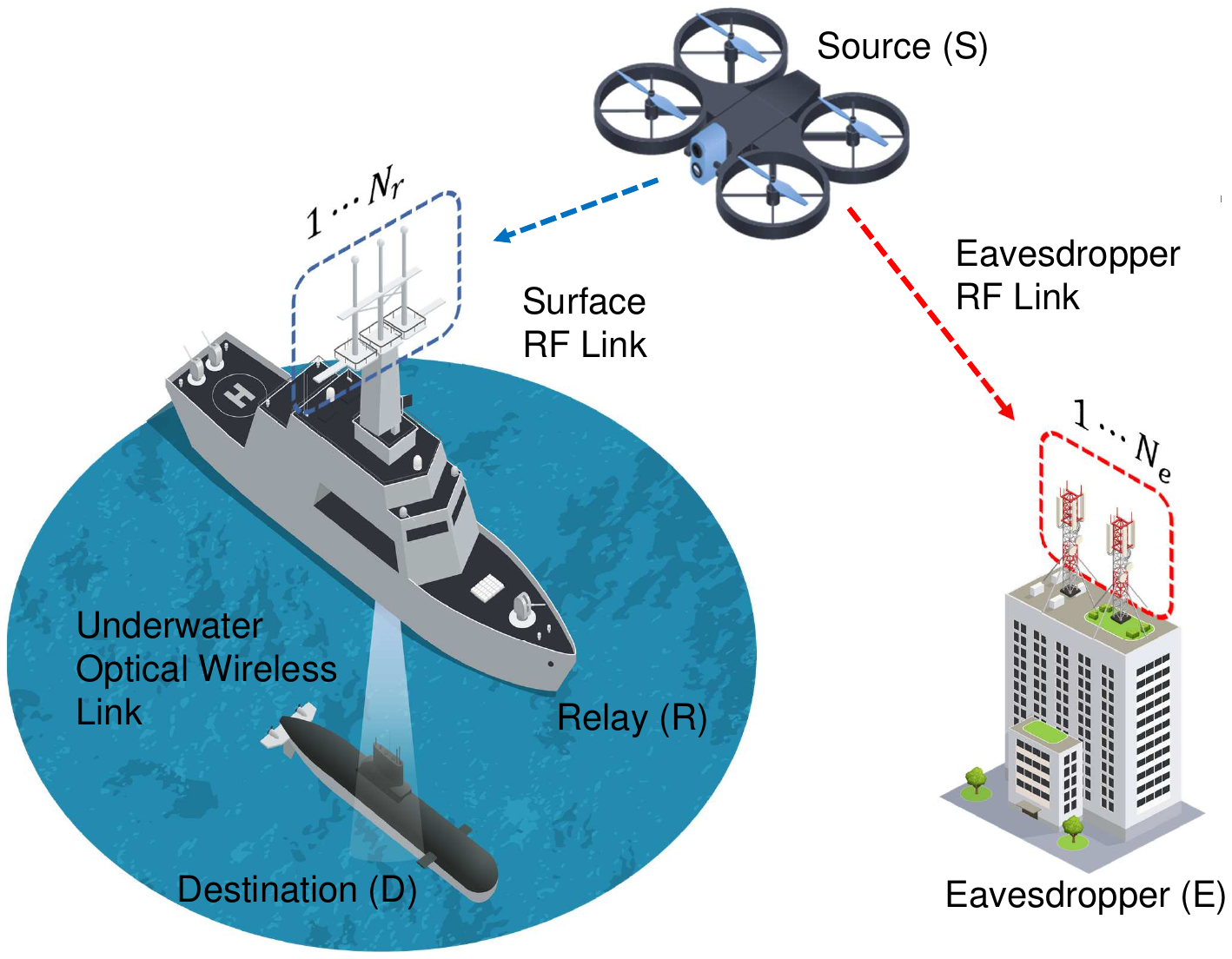}}
         \vspace{0mm }
    \caption{System model incorporating the source ($S$), relay ($R$), eavesdropper ($E$), and destination ($D$).}
       \label{f11}
   \end{figure*}
A four-node mixed RF-UOWC scenario is exhibited in Fig. \ref{f11}, where the total transmission process transpires into two-time slots since the typical RF signal strategy is not applicable for the UOWC. In the first time slot, a source node $S$ (e.g. ground control and monitor station, ship, airborne), transmits the concealed information to a relay node $R$ (e.g. floating buoy) via the RF link, whereas an unauthorized potential eavesdropper $E$ tries to theft away from the unrevealed stream of data from $S$ via utilizing the same RF link. To enhance the data speed, we assume all the three nodes (i.e. source, eavesdropper, and receiving end of the relay) are furnished with multiple antennas denoted as $N_{s}$, $N_{e}$, and $N_{r}$, respectively. In the proposed model, the TAS scheme specifies the strongest antenna that dispatches the signal with the highest SNR to $R$ while all other antennas remain unused thereby ensuring the utmost spectral efficiency. Besides transmit diversity at the source, we also assume the relay $R$ exploits the MRC technique to combine the received signals at $N_{r}$ antennas thereby ensuring the maximum instantaneous SNR at the relay.

Applying MRC system at $R$, the instantaneous SNR of $S-R$ link can be expressed as
\begin{align}
\gamma_{r}=\frac{T_{x}}{P_{R}\textbf{I}_{N_{R}}} \|\bold{h}_{s^{*},r}\|^{2},
\end{align}
where $T_{x}$ represents transmit power of the $x^{th}$ ($x=1,2\cdots N_{S}$) selected antenna, $\textbf{h}_{s^{*},r}$ denotes channel gain between the selected antenna at $S$ and $R$, $P_{R}$ symbolizes the noise powers at $R$, and $\textbf{I}_{N_R}$ is an identity matrix of order $N \times N$. We assume $E$ is unconscious about the TAS protocol\footnote{We assume passive eavesdropping scenario i.e. $S$ is unaware of CSI of the $S-E$ link. Hence, $E$ remains silent and $S$ transmits information at target secrecy rate. But $R$ is aware of CSI of the channels corresponding to each transmit antenna through the pilot sequence.}. Hence, implementing MRC technique at $E$, the instantaneous SNR of $S-E$ link can be written as  
\begin{align}
\gamma_{e}=\frac{T_{x}}{P_{E}\textbf{I}_{N_{E}}}\|\textbf{h}_{s^{*},e}\|^{2},
\end{align}
where $h_{s^{*},e}$ represents the channel gain between $S$ and $E$ and $P_{E}$ denotes the noise powers at $E$. In the second time slot, the stipulated variable gain AF relay receives the transmitted signal and transforms the received RF signal into optical form. Then, it amplifies and forwards this signal to the destination node $D$ (e.g. submarine). As a result, the instantaneous SNR of the $R-D$ link can be expressed as
\begin{align}
\gamma_{d}=\frac{T_{r}}{P_{D}}\|h_{r,d}\|^{2},
\end{align}
where $T_{r}$ denotes the optical power transmitted from R, $h_{r,d}$ is the channel gain of $R-D$ link, and $P_{D}$ signifies the optical noise imposed at D. It is mentioned the gain of relay operation can be considered as $\eta$ that is also incorporated within $h_{r,d}$ channel gain. Applying channel state information (CSI) assisted relaying technique at the relay, the received instantaneous SNR at $D$ for mixed RF-UOWC system can be expressed as \cite[Eq.~(1)]{9139494}
 \begin{align}
 \gamma_{f}&=\frac{\gamma_{r} \gamma _{d}}{\gamma_{r}+\gamma_{d}+1}
 \cong min\left\{\gamma_{r}, \gamma_{d}\right\}.
 \end{align}


\subsection{SNR of the RF Link}
We assume $\eta$-$\mu$ distribution for the $S$-$R$ RF link as it can perfectly characterize the practical non line of sight scenarios undergoing small scale fading \cite{yacoub2007kappa}. The probability density function (PDF) $\gamma_{r}$ exploiting the MRC technique is expressed as \cite[Eq.~(2),]{yang2018physical}
\begin{align}
\label{eqn:rfpdf1}
\nonumber
f_{\gamma_{r}}(x)&=\frac{2\sqrt{\pi}(N_{r}\mu_{r})^{N_{r}\mu_{r}+\frac{1}{2}}h_{r}^{N_{r}\mu_{r}}x^{N_{r}\mu_{r}-\frac{1}{2}}}{\Gamma(N_{r}\mu_{r})H_{r}^{N_{r}\mu_{r}-\frac{1}{2}}\varphi_{r}^{N_{r}\mu_{r}+\frac{1}{2}}}
\\
& \times  e^{-2N_{r}\mu_{r}h_{r}\frac{x}{\varphi_{r}}}I_{N_{r}\mu_{r}-\frac{1}{2}}\biggl(\frac{2N_{r}\mu_{r}H_{r}x}{\varphi_{r}}\biggl),
\end{align}
where $I_{v}(.)$ specifies the first kind modified bessel function with order $v$ \cite{watson1995treatise} and $\varphi_{r}$ denotes the average SNR of $S-R$ link. Here, $\eta$-$\mu$ distribution may appear in two distinct formats, for which two complementary physical scenarios can be elucidated. The two formats can be described in terms of $h_{r}$ and $H_{r}$ that are given as:

\noindent
\textbf{Format I:} $h_{r}=\frac{2+\eta_{r}^{-1}+\eta_{r}}{4}$ and $H_{r}=\frac{\eta_{r}^{-1}-\eta_{r}}{4}$ with $0<\eta_{r}<\infty$,

\noindent
\textbf{Format II:}
$h_{r}=\frac{1}{1-\eta^{2}_{r}}$ and $H_{r}=\frac{\eta_{r}}{1-\eta^{2}_{r}}$ with $-1<\eta_{r}<1$. 

\noindent
Here in Format I, $\eta_{r}$ signifies the scattered-wave power ratio between in phase and quadrature components within each clusters and in Format II, $\eta_{r}$ defines correlation between the powers of in-phase and quadrature scattered waves in each multi-path cluster \cite{yacoub2007kappa}.\footnote{It is important to mention Format II can be easily derived from Format I \cite{peppas2013performance}. Therefore, without any loss of generality we assume Format I for the $S-R$ link.} In both formats, $\mu_{r}$ denotes the multi-path clusters with $\mu_{r}>0$. However, determination of secrecy behaviour utilizing \eqref{eqn:rfpdf1} is significantly challenging. In such a case, most of the previous works are accomplished assuming $\mu_{r}$ to be an integer allowing for tractable evaluation wherein \eqref{eqn:rfpdf1} is expressed in an alternative form as
\begin{align}
\label{eqn:rfpdf2}
f_{\gamma_{r}}(x)&=\frac{h_{r}^{N_{r}\mu_{r}}}{H_{r}^{N_{r}\mu_{r}}\Gamma(N_{r}\mu_{r})}
\sum_{\alpha=1}^{2}\sum_{\beta=0}^{N_{r}\mu_{r}-1}B_{\alpha,\beta}x^{N_{r}\mu_{r}-\beta-1}e^{-\psi_\alpha x},
\end{align}
where $B_{1,\beta}=\frac{\Gamma(N_{r}\mu_{r}+\beta)(N_{r}\mu_{r})^{N_{r}\mu_{r}-\beta}(-1)^{\beta}}{\beta!\Gamma(N_{r}\mu_{r}-\beta)4^{\beta}\varphi_{r}^{N_{r}\mu_{r}-\beta}H_{r}^{\beta}}$, $B_{2,\beta}=\frac{\Gamma(N_{r}\mu_{r}+\beta)(N_{r}\mu_{r})^{N_{r}\mu_{r}-\beta}(-1)^{N_{r}\mu_{r}}}{\beta!\Gamma(N_{r}\mu_{r}-\beta)4^{\beta}\varphi_{r}^{N_{r}\mu_{r}-\beta}H_{r}^{\beta}}$, $\psi_{1}=\frac{2N_{r}\mu_{r}(h_{r}-H_{r})}{\varphi_{r}}$, and $\psi_{2}=\frac{2N_{r}\mu_{r}(h_{r}+H_{r})}{\varphi_{r}}$.
Now, utilizing the definition, cumulative density function (CDF) of $\gamma_{r}$ is written as
\begin{align}\label{eqn:rfcdf1}
F_{\gamma_{r}}(x)=\int_{0}^{x}f_{\gamma_{r}}(x) dx.
\end{align}
Replacing \eqref{eqn:rfpdf2} into \eqref{eqn:rfcdf1} and utilizing \cite[Eq.~(3.381.8)]{jeffrey2007table}, $F_{\gamma_{r}}(x)$ is obtained as
\begin{align}
\label{eqn:rfcdf2}
\nonumber
F_{\gamma_{r}}(x)&=1-\frac{h_{r}^{N_{r}\mu_{r}}}{H_{r}^{N_{r}\mu_{r}}\Gamma(N_{r}\mu_{r})}
\nonumber
\\
&\times \sum_{\alpha=1}^{2}\sum_{\beta=0}^{N_{r}\mu_{r}-1}\sum_{\phi=0}^{N_{r}\mu_{r}-\beta-1} L_{r}x^{\phi}e^{-\psi_\alpha x},
\end{align}
where 
$L_{r}=\frac{C_{\alpha,\beta}\psi_\alpha^{\phi}}{\phi!}$,
$C_{1,\beta}=\frac{(-1)^{\beta}\Gamma(N_{r}\mu_{r}+\beta)H_{r}^{-\beta}}{\beta!2^{N_{r}\mu_{r}+\beta}(h_{r}-H_{r})^{N_{r}\mu_{r}-\beta}}$, and
$C_{2,\beta}=\frac{(-1)^{N_{r}\mu_{r}}\Gamma(N_{r}\mu_{r}+\beta)H_{r}^{-\beta}}{\beta!2^{N_{r}\mu_{r}+\beta}(h_{r}+H_{r})^{N_{r}\mu_{r}-\beta}}$. Utilizing the order statistics for transmit antenna selection and assuming $\gamma_{r}*$ as an ordered variable, the CDF in \eqref{eqn:rfcdf1} is finally written as
\begin{align}
\label{eqn:antenna}
F_{\gamma_{r}*}(x)=[F_{\gamma_{r}}(x)]^{N_{S}}.
\end{align}
Substituting \eqref{eqn:rfcdf2} into \eqref{eqn:antenna} and performing some algebraic manipulations after applying multinomial theorem, we obtain
\begin{align}
\nonumber
\label{eqn:rfcdf222}
F_{\gamma_{r}*}(x)&=\sum_{m=0}^{N_{S}} \binom{N_{S}}{m}\sum_{n=0}^{m} \binom{m}{n}\sum_{u=0}^{(m-n)(N_{r}\mu_{r}-1)}\sum_{v=0}^{n(N_{r}\mu_{r}-1)}
\\
&\times Re^{-\varphi x}x^{u+v},
\end{align}
where $R=V^{m}\psi_{u}\psi_{v}$, $\varphi=(m-n)\psi_{1}+n\psi_{2}$, $V=-\frac{h_{r}^{N_{r}\mu_{r}}}{H_{r}^{N_{r}\mu_{r}}\Gamma(N_{r}\mu_{r})}$, 
and $\psi_{u}$, $\psi_{v}$ are expressed as the coefficients of $x^{u}$ and $x^{v}$ in the expansions of $\biggl(\sum_{\beta=0}^{N_{r}\mu_{r}-1}\sum_{\phi=0}^{N_{r}\mu_{r}-\beta-1}\frac{C_{1,\beta}}{\phi!}\psi_{1}^{\phi}x^{\phi}\biggl)^{m-n}$ and  $\biggl(\sum_{\beta=0}^{N_{r}\mu_{r}-1}\sum_{\phi=0}^{N_{r}\mu_{r}-\beta-1}\frac{C_{2,\beta}}{\phi!}\psi_{2}^{\phi}x^{\phi}\biggl)^{n}$, respectively. In addition, $\eta$-$\mu$ fading proposes generality that incorporates one-sided Gaussian, Nakagami-$m$, Hoyt, and Rayleigh distributions as special scenarios by selecting particular values of $\eta$ and $\mu$ \cite{yacoub2007kappa}.


\subsection{SNR of the UOWC link}
We consider mEGG model for the $R-D$ link for its generic characteristics. The PDF of $\gamma_{d}$ is expressed as \cite{li2021performance}
\textcolor{black}{\begin{align}\label{eqn:fsopdf}
f_{\gamma_{d}}(x)=\sum_{i=1}^{2}\mathcal{M}_{i}x^{-1}G_{1,2}^{2,0}\left[\mathcal{N}_{i}x^{\mathcal{V}_{i}}\biggl |
\begin{array}{c}
\mathcal{W}_{i} \\
\mathcal{U}_{i}, \mathcal{K}_{i} \\
\end{array}
\right],
\end{align}
where $\mathcal{N}_{1}=\frac{1}{\lambda A_{o}\Psi_{r}^{\frac{1}{r}}}$, $\mathcal{V}_{1}=\frac{1}{r}$, $\mathcal{U}_{1}=1$, $\mathcal{M}_{1}=\frac{\omega\xi^{2}}{r}$, $\mathcal{W}_{1}=\xi^{2}+1$, $\mathcal{K}_{1}=\xi^{2}$, $\mathcal{N}_{2}=\frac{1}{A_{o}^{c}b^{c}\Psi_{r}^{\frac{c}{r}}}$, $\mathcal{V}_{2}=\frac{c}{r}$, $\mathcal{U}_{2}=a$, $\mathcal{M}_{2}=\frac{\xi^{2}(1-\omega)}{r\Gamma(a)}$, $\mathcal{W}_{2}=\frac{\xi^{2}}{c}+1$, $\mathcal{K}_{2}=\frac{\xi^{2}}{c}$,  $\lambda$ defines the exponentially distribution parameter, $a$, $b$, and $c$ stand for three GG distribution parameters, $\omega$ denotes the mixture weight with $0<\omega<1$, and $\xi$ denotes the pointing error wheres $A_{o}$ is the constant related to $\xi$ \cite{li2021performance}. Here, $\varphi_{d}$ is the average SNR of $R-D$ link that is related to electrical SNR denoted as $\Psi_{r}$ as defined in \cite{9466489},} where $r$ denotes the detection technique (i.e. $r=1$ specifies heterodyne detection (HD) technique and $r=2$ signifies intensity modulation / direct detection (IM/DD) technique). In \cite{zedini2019unified}, the values of $\omega$, $\lambda$, $a$, $b$ and, $c$ depending upon various UWT (i.e. weak to severe) influenced by different level of air bubbles, temperature gradients, and water salinity are determined experimentally that reveals a strong agreement with the measured data.\footnote{It is noteworthy that for the case with $c=1$, PDF of mEGG model takes the form of EG model \cite{badrudduza2021security}. In addition, mEGG model is also equivalent to generalized Gamma (GG) and exponential distributions \cite{zedini2019unified}.} \cite[Table~I]{zedini2019unified} introduces simultaneous impact of air bubbles corresponding to temperature gradients that gradually generates weak, moderate, and severe turbulence conditions. In addition, increase in air bubbles level and/or water temperature gradients strengthens the UWT thereby raises the scintillation index greatly. \cite[Table~II]{zedini2019unified} presents various turbulence scenarios along with level of air bubbles in both fresh and salty water environments considering a thermally uniform UOWC link.

By substituting \eqref{eqn:fsopdf} into \eqref{eqn:rfcdf1}, the CDF for UOWC link is obtained as
\textcolor{black}{\begin{align}\label{eqn:fsocdf}
F_{\gamma_{d}}(x)=\sum_{i=1}^{2}\mathcal{S}_{i}G_{2,3}^{2,1}\left[\mathcal{N}_{i}x^{\mathcal{V}_{i}}\biggl |
\begin{array}{c}
 1, \mathcal{W}_{i} \\
 \mathcal{U}_{i}, \mathcal{K}_{i}, 0 \\
\end{array}
\right],
\end{align}
where $\mathcal{S}_{1}=\omega\xi^{2}$ and $\mathcal{S}_{2}=\frac{\xi^{2}(1-\omega)}{c\Gamma(a)}$.}

\subsection{SNR of the Eavesdropper link}
The eavesdropper channel (i.e. $S-E$ link) also experiences $\eta-\mu$ fading distribution, and similar to $S-R$ RF link, the PDF of $\gamma_{e}$ is expressed as \cite[Eq.~(3)]{yang2018physical}
\begin{align}
\label{eqn:eavpdf}
f_{\gamma_{e}}(x)&=\frac{h_{e}^{N_{e}\mu_{e}}}{H_{e}^{N_{e}\mu_{e}}\Gamma(N_{e}\mu_{e})}
\sum_{\delta=1}^{2}\sum_{\theta=0}^{N_{e}\mu_{e}-1}D_{\delta,\theta}x^{N_{e}\mu_{e}-\theta-1}e^{-\Omega_{\delta}x},
\end{align}
where $D_{1,\theta}=\frac{\Gamma(N_{e}\mu_{e}+\theta)(N_{e}\mu_{e})^{N_{e}\mu_{e}-\theta}(-1)^{\theta}}{\theta!\Gamma(N_{e}\mu_{e}-\theta)4^{\theta}\varphi_{e}^{N_{e}\mu_{e}-\theta}H_{e}^{\theta}}$, $D_{2,\theta}=\frac{\Gamma(N_{e}\mu_{e}+\theta)(N_{e}\mu_{e})^{N_{e}\mu_{e}-\theta}(-1)^{N_{e}\mu_{e}}}{\theta!\Gamma(N_{e}\mu_{e}-\theta)4^{\theta}\varphi_{e}^{N_{e}\mu_{e}-\theta}H_{e}^{\theta}}$,
$\Omega_{1}=\frac{2N_{e}\mu_{e}(h_{e}-H_{e})}{\varphi_{e}}$, $\Omega_{2}=\frac{2N_{e}\mu_{e}(h_{e}+H_{e})}{\varphi_{e}}$, and  $\varphi_{e}$ symbolizes the average SNR of $S-E$ link. According to the Format I, $h_{e}=\frac{2+\eta_{e}^{-1}+\eta_{e}}{4}$, $H_{e}=\frac{\eta_{e}^{-1}-\eta_{e}}{4}$, and $\eta_{e}$ as well as $\mu_{e}$ of $S-E$ link have same definitions as in $S-R$ link. According to \eqref{eqn:rfcdf2}, making use of \eqref{eqn:eavpdf}, $F_{\gamma_{e}}$ is obtained as
\begin{align}
\label{eqn:eavcdf}
\nonumber
F_{\gamma_{e}}(x)&=1-\frac{h_{e}^{N_{e}\mu_{e}}}{H_{e}^{N_{e}\mu_{e}}\Gamma(N_{e}\mu_{e})}
\sum_{\delta=1}^{2}\sum_{\theta=0}^{N_{e}\mu_{e}-1}\sum_{\varsigma=0}^{N_{e}\mu_{e}-\theta-1}
\\
&\times L_{e}x^{\varsigma}e^{-\Omega_{\delta}x},
\end{align}
where $L_{e}=\frac{E_{\delta,\theta}\Omega_{\delta}^{\varsigma}}{\varsigma!}$, $E_{1,\theta}=\frac{(-1)^{\theta}\Gamma(N_{e}\mu_{e}+\theta)H_{e}^{-\theta}}{\theta!2^{N_{e}\mu_{e}+\theta}(h_{e}-H_{e})^{N_{e}\mu_{e}-\theta}}$, and $E_{2,\theta}=\frac{(-1)^{N_{e}\mu_{e}}\Gamma(N_{e}\mu_{e}+\theta)H_{e}^{-\theta}}{\theta!2^{N_{e}\mu_{e}+\theta}(h_{e}+H_{e})^{N_{e}\mu_{e}-\theta}}$.

However, utilizing \cite[Eqs.~(3.351.1) and (8.354.1)]{jeffrey2007table} and following \eqref{eqn:rfcdf1}, the CDF of $\gamma_{e}$ is calculated in an alternative form as
\begin{align}
\label{eqn:eavcdfnew}
\nonumber
F_{\gamma_{e}}(x)&= \frac{h_{e}^{N_{e}\mu_{e}}}{H_{e}^{N_{e}\mu_{e}}\Gamma(N_{e}\mu_{e})}
\sum_{\delta=1}^{2}\sum_{\theta=0}^{N_{e}\mu_{e}-1}\sum_{z_{1}=0}^{\infty} \frac{(-1)^{z_{1}}\Omega_{\delta}^{z_{1}}}{z_{1}!w_{z}}
\\
& \times D_{\delta,\theta} x^{w_{z}},
\end{align}
where $w_{z}=N_{e}\mu_{e}-\theta+z_{1}$. Note that \eqref{eqn:eavcdfnew} is in the form of infinite series, it converge rapidly and steadily to accurate results after twenty terms \cite{jameson2016incomplete}.


\subsection{SNR of Dual hop RF-UOWC LINK}
In this section, analytical expression of CDF for end-to-end SNR of the considered mixed RF-UOWC system is demonstrated. Since variable AF gain relaying strategy is implemented at the relay node, CDF of $\gamma_{f}$ is obtained as \cite[Eq.~(15)]{odeyemi2019impactdualhop} 
\begin{align}
\label{eqn:dualhopcdf05}
\nonumber
F_{\gamma_{f}}(x)&\approx \mathrm{Pr}\left\{\mathrm{min}(\gamma_{r}, \gamma _{d})<\gamma_{f}\right\}
\\
\nonumber
&= 1-\mathrm{Pr}[\gamma_{r}>\gamma_{f}] \mathrm{Pr}[\gamma_{d}>\gamma_{f}]
\\
\nonumber
&= 1- (1- F_{\gamma_{r}}(x))(1- F_{\gamma_{d}}(x))
\\
&= F_{\gamma_{r}}(x)+F_{\gamma_{d}}(x)-F_{\gamma_{r}}(x)F_{\gamma_{d}}(x).
\end{align}
Now, substituting \eqref{eqn:rfcdf222} and \eqref{eqn:fsocdf} into \eqref{eqn:dualhopcdf05} along with some simplifications and  algebraic manipulations, the CDF of $\gamma_{f}$ is obtained as
\textcolor{black}{\begin{align}
\label{eqn:dualhopcdf01}
\nonumber
F_{\gamma_{f}}(x)&=\sum_{m=0}^{N_{S}} \binom{N_{S}}{m}\sum_{n=0}^{m} \binom{m}{n}\sum_{u=0}^{(m-n)(N_{r}\mu_{r}-1)}\sum_{v=0}^{n(N_{r}\mu_{r}-1)}
\\
\nonumber
&\times R\biggl(1-\sum_{i=1}^{2}\mathcal{S}_{i}G_{2,3}^{2,1}\left[\mathcal{N}_{i}x^{\mathcal{V}_{i}}\biggl |
\begin{array}{c}
 1, \mathcal{W}_{i} \\
 \mathcal{U}_{i}, \mathcal{K}_{i}, 0 \\
\end{array}
\right]\biggl)
\\
&\times e^{-\varphi x}x^{u+v}+\sum_{i=1}^{2}\mathcal{S}_{i}G_{2,3}^{2,1}\left[\mathcal{N}_{i}x^{\mathcal{V}_{i}}\biggl |
\begin{array}{c}
 1, \mathcal{W}_{i} \\
 \mathcal{U}_{i}, \mathcal{K}_{i}, 0 \\
\end{array}
\right].
\end{align}}
To the best of authors' knowledge, the developed CDF expression in \eqref{eqn:dualhopcdf01} for mixed RF-UOWC systems is unique as the combination of $\eta$-$\mu$ and mEGG model while considering TAS/MRC approach are not reported in any previous works. Moreover, both $\eta$-$\mu$ as well as mEGG channels are generalized in nature that provides advantages of modeling several existing scenarios as special cases.

\section{{\color{black}Outage Performance Analysis}}
{\color{black}This section demonstrates outage performance ($OP$) of the proposed model in absence of any eavesdroppers using CDF of end-to-end dual-hop SNR as derived in \eqref{eqn:dualhopcdf01}. $OP$ is defined as the probability of $\gamma_{f}$ falls below a prefixed threshold $\gamma_{th}$. Mathematically, $OP$ is the CDF of $\gamma_{f}$ evaluated at $\gamma_{th}$. Hence, $OP$ can be calculated from \eqref{eqn:dualhopcdf01} as
\begin{align}
    OP=\Pr(\gamma_{f}<\gamma_{th})=F_{\gamma_{f}}(\gamma_{th}).
\end{align}
}

\section{Secrecy Performance Analysis}
To investigate the PLS, three important performance metrics (such as ASC, SOP, and probability of SPSC) are demonstrated in this section utilizing \eqref{eqn:eavpdf}, \eqref{eqn:eavcdf}, \eqref{eqn:eavcdfnew}, and \eqref{eqn:dualhopcdf01}.

\subsection{Average Secrecy Capacity Analysis}
ASC is one of the most significant performance metrics to investigate secrecy characteristics of the considered RF-UOWC system. ASC can be explained as the average of instantaneous secrecy rate \cite{9427251}. Mathematically, ASC is defined as \cite[Eq.~(15)]{[45]wang2014}
\begin{align}\label{eqn:asc}
ASC=\int_{0}^{\infty}\frac{F_{\gamma_{e}}(x)}{1+x}\left\{1-F_{\gamma_{f}}(x)\right\}dx.
\end{align}
By placing \eqref{eqn:eavcdf} and \eqref{eqn:dualhopcdf01} into \eqref{eqn:asc}, analytical expression of ASC is derived as \eqref{eqn:asc1}, where $\mathcal{A}_{1}$, $\mathcal{A}_{2}$, $\mathcal{A}_{3}$, $\mathcal{A}_{4}$, $\mathcal{A}_{5}$, $\mathcal{A}_{6}$, $\mathcal{A}_{7}$, and $\mathcal{A}_{8}$ are the eight integral terms determined as follows.
\begin{figure*}[!t]
\textcolor{black}{\begin{align}
\label{eqn:asc1}
\nonumber
ASC_{1}&=\mathcal{A}_{1}-\frac{h_{e}^{N_{e}\mu_{e}}}{H_{e}^{N_{e}\mu_{e}}\Gamma(N_{e}\mu_{e})}\sum_{\delta=1}^{2}\sum_{\theta=0}^{N_{e}\mu_{e}-1}\sum_{\varsigma=0}^{N_{e}\mu_{e}-\theta-1}L_{e}\biggl(\mathcal{A}_{2}-\sum_{i=1}^{2}\mathcal{S}_{i}\mathcal{A}_{6}\biggl)-\sum_{i=1}^{2}\mathcal{S}_{i}\mathcal{A}_{5}- \sum_{m=0}^{N_{S}} \binom{N_{S}}{m}\sum_{n=0}^{m} \binom{m}{n}
\\
&\times \sum_{u=0}^{(m-n)(N_{r}\mu_{r}-1)}\sum_{v=0}^{n(N_{r}\mu_{r}-1)}R\biggl[\mathcal{A}_{3}-\frac{h_{e}^{N_{e}\mu_{e}}}{H_{e}^{N_{e}\mu_{e}}\Gamma(N_{e}\mu_{e})}\sum_{\delta=1}^{2}\sum_{\theta=0}^{N_{e}\mu_{e}-1}\sum_{\varsigma=0}^{N_{e}\mu_{e}-\theta-1}L_{e}\biggl(\mathcal{A}_{4}-\sum_{i=1}^{2}\mathcal{S}_{i}\mathcal{A}_{8}\biggl)-\sum_{i=1}^{2}\mathcal{S}_{i}\mathcal{A}_{7}\biggl].
\end{align}}
\hrulefill
\end{figure*}
$\mathcal{A}_{1}$ is derived as 
\begin{align}
\label{aaa}
    \mathcal{A}_{1}&=\int_{0}^{\infty}\frac{dx}{1+x}=\ln(1+z),
\end{align} 
where $z$ is a very high value ( $ \rightarrow \infty $ ). It can be noted that although the integration in \eqref{aaa} can not be performed, we can easily obtain values via numerical approach.
With the help of identities \cite[Eqs.~(8.4.2.5) and (8.4.3.1)]{Calculationbook02} for mathematical conversion of terms $\frac{1}{1+x}$ and $e^{-\Omega_{\delta}x}$ into Meijer's $G$ function and performing integration following \cite[Eq.~(7.811.1)]{jeffrey2007table}, $\mathcal{A}_{2}$ is obtained as
\begin{align}
\nonumber
\mathcal{A}_{2}&=\int_{0}^{\infty}\frac{x^{\varsigma}}{1+x}e^{-\Omega_{\delta}x} dx
\\
\nonumber
&=\int_{0}^{\infty}x^{\varsigma}G_{1,1}^{1,1}\left[x\biggl |
\begin{array}{c}
0 \\
0 \\
\end{array}
\right]G_{0,1}^{1,0}\left[\Omega_{\delta}x\biggl |
\begin{array}{c}
- \\
0 \\
\end{array}
\right] dx
\\
\nonumber
&=\int_{0}^{\infty}G_{1,1}^{1,1}\left[x\biggl |
\begin{array}{c}
\varsigma \\
\varsigma \\
\end{array}
\right]G_{0,1}^{1,0}\left[\Omega_{\delta}x\biggl |
\begin{array}{c}
- \\
0 \\
\end{array}
\right] dx
\\
&=G_{1,2}^{2,1}\left[\Omega_{\delta}\biggl |
\begin{array}{c}
-\varsigma \\
0, -\varsigma \\
\end{array}
\right].
\end{align}
Using similar identities as were utilized for $\mathcal{A}_{2}$, $\mathcal{A}_{3}$ is derived as
\begin{align}
\nonumber
\mathcal{A}_{3}&=\int_{0}^{\infty}\frac{x^{u+v}}{1+x}e^{-\varphi x} dx
\\
\nonumber
&=\int_{0}^{\infty}x^{u+v}G_{1,1}^{1,1}\left[x\biggl |
\begin{array}{c}
0 \\
0 \\
\end{array}
\right]G_{0,1}^{1,0}\left[\varphi x\biggl |
\begin{array}{c}
- \\
0 \\
\end{array}
\right] dx
\\
\nonumber
&=\int_{0}^{\infty}G_{1,1}^{1,1}\left[x\biggl |
\begin{array}{c}
u+v \\
u+v \\
\end{array}
\right]G_{0,1}^{1,0}\left[\varphi x\biggl |
\begin{array}{c}
- \\
0 \\
\end{array}
\right] dx
\\
&=G_{1,2}^{2,1}\left[\varphi\biggl |
\begin{array}{c}
-(u+v) \\
0, -(u+v) \\
\end{array}
\right].
\end{align}
By following similar approach of $\mathcal{A}_{2}$ and $\mathcal{A}_{3}$, $\mathcal{A}_{4}$ is obtained as
\begin{align}
\nonumber
\mathcal{A}_{4}&=\int_{0}^{\infty}\frac{x^{u+v+\varsigma}}{1+x}e^{-(\Omega_{\delta}+\varphi) x} dx
\\
\nonumber
&=\int_{0}^{\infty}x^{u+v+\varsigma}G_{1,1}^{1,1}\left[x\biggl |
\begin{array}{c}
0 \\
0 \\
\end{array}
\right]
G_{0,1}^{1,0}\left[(\Omega_{\delta}+\varphi) x\biggl |
\begin{array}{c}
- \\
0 \\
\end{array}
\right] dx
\\
\nonumber
&=\int_{0}^{\infty}G_{1,1}^{1,1}\left[x\biggl |
\begin{array}{c}
u+v+\varsigma \\
u+v+\varsigma \\
\end{array}
\right]
G_{0,1}^{1,0}\left[(\Omega_{\delta}+\varphi) x\biggl |
\begin{array}{c}
- \\
0 \\
\end{array}
\right] dx
\\
&=G_{1,2}^{2,1}\left[(\Omega_{\delta}+\varphi)\biggl |
\begin{array}{c}
-(u+v+\varsigma) \\
0, -(u+v+\varsigma) \\
\end{array}
\right].
\end{align}
Similar to $\mathcal{A}_{2}$, converting $\frac{1}{1+x}$ into Meijer's $G$ function and performing integration with the help of \cite[Eq.~(2.24.1.1)]{Calculationbook02}, $\mathcal{A}_{5}$ is derived as
\textcolor{black}{\begin{align}
\nonumber
\mathcal{A}_{5}&=\int_{0}^{\infty}\frac{1}{1+x}G_{2,3}^{2,1}\left[\mathcal{N}_{i}x^{\mathcal{V}_{i}}\biggl |
\begin{array}{c}
 1, \mathcal{W}_{i} \\
 \mathcal{U}_{i}, \mathcal{K}_{i}, 0 \\
\end{array}
\right]dx
\\
\nonumber
&=\int_{0}^{\infty}G_{1,1}^{1,1}\left[x\biggl |
\begin{array}{c}
0 \\
0 \\
\end{array}
\right]G_{2,3}^{2,1}\left[\mathcal{N}_{i}x^{\mathcal{V}_{i}}\biggl |
\begin{array}{c}
 1, \mathcal{W}_{i} \\
 \mathcal{U}_{i}, \mathcal{K}_{i}, 0 \\
\end{array}
\right] dx
\\
\nonumber
&=\frac{1}{2\pi^{(\mathcal{V}_{i}-1)}} 
\\
&\times G_{2+\mathcal{V}_{i},3+\mathcal{V}_{i}}^{2+\mathcal{V}_{i},1+\mathcal{V}_{i}}\left[\mathcal{N}_{i}\biggl |
\begin{array}{c}
 1, \Delta(\mathcal{V}_{i},0), \Delta(1, \mathcal{W}_{i}) \\
\Delta(1, \mathcal{U}_{i}), \Delta(1, \mathcal{K}_{i}),  \Delta(\mathcal{V}_{i},0), 0 \\
\end{array}
\right].
\end{align}}
Utilizing the integral identities \cite[Eqs.~(8.4.2.5) and (8.4.3.1)]{Calculationbook02}, $\mathcal{A}_{6}$ is obtained as
\textcolor{black}{\begin{align}
\nonumber
\mathcal{A}_{6}&=\int_{0}^{\infty}\frac{x^{\varsigma}}{1+x}e^{-\Omega_{\delta}x}G_{2,3}^{2,1}\left[\mathcal{N}_{i}x^{\mathcal{V}_{i}}\biggl |
\begin{array}{c}
 1, \mathcal{W}_{i} \\
 \mathcal{U}_{i}, \mathcal{K}_{i}, 0 \\
\end{array}
\right] dx
\\
\nonumber
&=\int_{0}^{\infty}x^{\varsigma}G_{1,1}^{1,1}\left[x\biggl |
\begin{array}{c}
0 \\
0 \\
\end{array}
\right]G_{0,1}^{1,0}\left[\Omega_{\delta}x\biggl |
\begin{array}{c}
- \\
0 \\
\end{array}
\right] 
\\
\nonumber
& \times G_{2,3}^{2,1}\left[\mathcal{N}_{i}x^{\mathcal{V}_{i}}\biggl |
\begin{array}{c}
 1, \mathcal{W}_{i} \\
 \mathcal{U}_{i}, \mathcal{K}_{i}, 0 \\
\end{array}
\right]dx.
\end{align}}
Now, for tractable analysis, transforming Meijer's $G$ functions into Fox's $H$ functions utilizing \cite[Eq.~(6.2.8)]{springer1979algebra} and performing integration by means of \cite[Eq.~(2.3)]{mittal1972integral} and \cite[Eq.~(3)]{lei2017secrecy}, $\mathcal{A}_{6}$ is obtained as
\textcolor{black}{\begin{align}
\nonumber
&\mathcal{A}_{6}=\int_{0}^{\infty}x^{\varsigma}H_{1,1}^{1,1}\left[x\biggl |
\begin{array}{c}
(0, 1) \\
(0, 1) \\
\end{array}
\right]H_{0,1}^{1,0}\left[\Omega_{\delta}x\biggl |
\begin{array}{c}
- \\
(0, 1) \\
\end{array}
\right] 
\\
\nonumber
&\times H_{2,3}^{2,1}\left[\mathcal{N}_{i}x^{\mathcal{V}_{i}}\biggl |
\begin{array}{c}
 (1, 1), (\mathcal{W}_{i}, 1) \\
 (\mathcal{U}_{i}, 1), (\mathcal{K}_{i}, 1), (0, 1) \\
\end{array}
\right]dx
\\
\nonumber
&=\Omega_{\delta}^{-(\varsigma+1)}
\\
&\times H_{1,0:1,1:2,3}^{1,0:1,1:2,1}\biggl[\begin{array}{c}
J_{1}\\
J_{2}\\
\end{array}\biggl | \begin{array}{c}
(0, 1)\\
(0, 1)\\
\end{array}\biggl | \begin{array}{c}
(1,1), (\mathcal{W}_{i}, 1)\\
(\mathcal{U}_{i}, 1), (\mathcal{K}_{i}, 1), (0, 1)\\
\end{array}\biggl | J_{3},J_{4}\biggl],
\end{align}
where $J_{1}=(-\varsigma;1,\mathcal{V}_{i})$, $J_{2}=(1;-)$, $J_{3}=\frac{1}{\Omega_{\delta}}$, $J_{4}=\frac{\mathcal{N}_{i}}{\Omega_{\delta}^{\mathcal{V}_{i}}}$,} $H_{p,q}^{m,n}[.]$ is the Fox's $H$ function from \cite[Eq.~(1.2)]{mathai2009h}, and $H_{c_{1},d_{1}:c_{2},d_{2}:c_{3},d_{3}}^{x_{1},y_{1}:x_{2},y_{2}:x_{3},y_{3}}[.]$ is termed as the extended generalized bivariate Fox’s $H$ function (EGBFHF) explained in \cite[Eq.~(2.57)]{mathai2009h}.
Approaching with similar steps as described for $\mathcal{A}_{6}$, $\mathcal{A}_{7}$ is determined as 
\textcolor{black}{\begin{align}
\nonumber
&\mathcal{A}_{7}=\int_{0}^{\infty}\frac{x^{u+v}}{1+x}e^{-\varphi x} G_{2,3}^{2,1}\left[\mathcal{N}_{i}x^{\mathcal{V}_{i}}\biggl |
\begin{array}{c}
 1, \mathcal{W}_{i} \\
 \mathcal{U}_{i}, \mathcal{K}_{i}, 0 \\
\end{array}
\right] dx
\\
\nonumber
&=\int_{0}^{\infty}x^{u+v}H_{1,1}^{1,1}\left[x\biggl |
\begin{array}{c}
(0, 1) \\
(0, 1) \\
\end{array}
\right]H_{0,1}^{1,0}\left[\varphi x\biggl |
\begin{array}{c}
- \\
(0, 1) \\
\end{array}
\right] 
\\
\nonumber
&\times H_{2,3}^{2,1}\left[\mathcal{N}_{i}x^{\mathcal{V}_{i}}\biggl |
\begin{array}{c}
 (1, 1), (\mathcal{W}_{i}, 1) \\
 (\mathcal{U}_{i}, 1), (\mathcal{K}_{i}, 1), (0, 1) \\
\end{array}
\right]dx
\\
\nonumber
&=\varphi^{-(u+v+1)}
\\
&\times H_{1,0:1,1:2,3}^{1,0:1,1:2,1}\biggl[\begin{array}{c}
J_{5}\\
J_{2}\\
\end{array}\biggl | \begin{array}{c}
(0,1)\\
(0,1)\\
\end{array}\biggl | \begin{array}{c}
(1,1), (\mathcal{W}_{i}, 1)\\
(\mathcal{U}_{i}, 1), (\mathcal{K}_{i}, 1), (0, 1)\\
\end{array}\biggl | J_{6},J_{7}\biggl],
\end{align}
where $J_{5}=(-u-v;1,\mathcal{V}_{i})$, $J_{6}=\frac{1}{\varphi}$, and $J_{7}=\frac{\mathcal{N}_{i}}{\varphi^{\mathcal{V}_{i}}}$.}
With some simple mathematical manipulations and according to $\mathcal{A}_{6}$ and $\mathcal{A}_{7}$, $\mathcal{A}_{8}$ is derived as
\textcolor{black}{\begin{align}
\nonumber
&\mathcal{A}_{8}=\int_{0}^{\infty}\frac{x^{u+v+\varsigma}}{1+x}e^{-(\Omega_{\delta}+\varphi) x}
G_{2,3}^{2,1}\left[\mathcal{N}_{i}x^{\mathcal{V}_{i}}\biggl |
\begin{array}{c}
 1, \mathcal{W}_{i} \\
 \mathcal{U}_{i}, \mathcal{K}_{i}, 0 \\
\end{array}
\right] dx
\\
\nonumber
&=\int_{0}^{\infty}x^{u+v+\varsigma}H_{1,1}^{1,1}\left[x\biggl |
\begin{array}{c}
(0, 1) \\
(0, 1) \\
\end{array}
\right]
\\
\nonumber
&\times H_{0,1}^{1,0}\left[(\Omega_{\delta}+\varphi) x\biggl |
\begin{array}{c}
- \\
(0, 1) \\
\end{array}
\right]  
\\
\nonumber 
&\times H_{2,3}^{2,1}\left[\mathcal{N}_{i}x^{\mathcal{V}_{i}}\biggl |
\begin{array}{c}
 (1, 1), (\mathcal{W}_{i}, 1) \\
 (\mathcal{U}_{i}, 1), (\mathcal{K}_{i}, 1), (0, 1) \\
\end{array}
\right] dx
\\
\nonumber
&=(\Omega_{\delta}+\varphi)^{-(u+v+\varsigma+1)}
\\
&\times H_{1,0:1,1:2,3}^{1,0:1,1:2,1}\biggl[\begin{array}{c}
J_{8}\\
J_{2}\\
\end{array}\biggl | \begin{array}{c}
(0,1)\\
(0,1)\\
\end{array}\biggl | \begin{array}{c}
(1,1), (\mathcal{W}_{i}, 1)\\
(\mathcal{U}_{i}, 1), (\mathcal{K}_{i}, 1), (0, 1)\\
\end{array}\biggl | J_{9},J_{10}\biggl],
\end{align}
where $J_{8}=(-u-v-\varsigma;1,\mathcal{V}_{i})$, $J_{9}=\frac{1}{\Omega_{\delta}+\varphi}$, and $J_{10}=\frac{\mathcal{N}_{i}}{(\Omega_{\delta}+\varphi)^{\mathcal{V}_{i}}}$.}

\begin{figure*}[!t]
\textcolor{black}{\begin{align}
\label{eqn:asc2}
\nonumber
ASC_{2}&=\frac{h_{e}^{N_{e}\mu_{e}}}{H_{e}^{N_{e}\mu_{e}}\Gamma(N_{e}\mu_{e})}
\sum_{\delta=1}^{2}\sum_{\theta=0}^{N_{e}\mu_{e}-1}\sum_{z_{1}=0}^{\infty} \frac{(-1)^{z_{1}}\Omega_{\delta}^{z_{1}}}{z_{1}!w_{z}}D_{\delta,\theta}\biggl[\mathcal{B}_{1}- \sum_{m=0}^{N_{S}} \binom{N_{S}}{m}
\\
&\times\sum_{n=0}^{m} \binom{m}{n}\sum_{u=0}^{(m-n)(N_{r}\mu_{r}-1)}\sum_{v=0}^{n(N_{r}\mu_{r}-1)} R\biggl(\mathcal{B}_{2}-\sum_{i=1}^{2}\mathcal{S}_{i}\mathcal{B}_{3}\biggl)-\sum_{i=1}^{2}\mathcal{S}_{i}\mathcal{B}_{4}\biggl].
\end{align}}
\hrulefill
\end{figure*}

Alternatively, the analytical expression of ASC\footnote{Note that in ASC$_{1}$, the final expression of $\mathcal{A}_{1}$ can not be obtained in closed-form, and hence we demonstrate numerical analysis to obtain the analytical results corresponding to this particular term. We also demonstrate similar analytical results with ASC$_{2}$ by eliminating the aforementioned integration complexity in terms of infinite series representation of lower incomplete Gamma function while obtaining the CDF of eavesdropper channel's SNR shown in \eqref{eqn:eavcdfnew}.} is also derived utilizing \eqref{eqn:eavcdfnew} and \eqref{eqn:dualhopcdf01}. The derived closed-form of this alternative ASC is expressed in \eqref{eqn:asc2}, where $\mathcal{B}_{1}$, $\mathcal{B}_{2}$, $\mathcal{B}_{3}$, and $\mathcal{B}_{4}$ are the four integral terms determined as follows.

Utilizing \cite[Eq.(3.194.3)]{jeffrey2007table}, $\mathcal{B}_{1}$ is obtained as
\begin{align}
\mathcal{B}_{1}&= \int_{0}^{\infty}\frac{x^{w_{z}}}{1+x}dx=B(w_{z}+1,-w_{z}),
\end{align}
where $B(x,y)$ denotes the well known beta function as defined in \cite{jeffrey2007table}.
Similar to $\mathcal{A}_{4}$, $\mathcal{B}_{2}$ is derived as
\begin{align}
\nonumber
\mathcal{B}_{2}&=\int_{0}^{\infty}\frac{x^{u+v+w_{z}}}{1+x}e^{-\varphi x} dx
\\
\nonumber
&=\int_{0}^{\infty}x^{u+v+w_{z}}G_{1,1}^{1,1}\left[x\biggl |
\begin{array}{c}
0 \\
0 \\
\end{array}
\right]
G_{0,1}^{1,0}\left[\varphi x\biggl |
\begin{array}{c}
- \\
0 \\
\end{array}
\right] dx
\\
\nonumber
&=\int_{0}^{\infty}G_{1,1}^{1,1}\left[x\biggl |
\begin{array}{c}
u+v+w_{z} \\
u+v+w_{z} \\
\end{array}
\right]
G_{0,1}^{1,0}\left[\varphi x\biggl |
\begin{array}{c}
- \\
0 \\
\end{array}
\right] dx
\\
&=G_{1,2}^{2,1}\left[\varphi \biggl |
\begin{array}{c}
-(u+v+w_{z}) \\
0, -(u+v+w_{z}) \\
\end{array}
\right].
\end{align}
Approaching with similar steps as described for $\mathcal{A}_{7}$, $\mathcal{B}_{3}$ is determined as 
\textcolor{black}{\begin{align}
\nonumber
&\mathcal{B}_{3}=\int_{0}^{\infty}\frac{x^{u+v+w_{z}}}{1+x}e^{-\varphi x} G_{2,3}^{2,1}\left[\mathcal{N}_{i}x^{\mathcal{V}_{i}}\biggl |
\begin{array}{c}
 1, \mathcal{W}_{i} \\
 \mathcal{U}_{i}, \mathcal{K}_{i}, 0 \\
\end{array}
\right] dx
\\
\nonumber
&=\int_{0}^{\infty}x^{u+v+w_{z}}H_{1,1}^{1,1}\left[x\biggl |
\begin{array}{c}
(0, 1) \\
(0, 1) \\
\end{array}
\right]H_{0,1}^{1,0}\left[\varphi x\biggl |
\begin{array}{c}
- \\
(0, 1) \\
\end{array}
\right] 
\\
\nonumber
&\times H_{2,3}^{2,1}\left[\mathcal{N}_{i}x^{\mathcal{V}_{i}}\biggl |
\begin{array}{c}
 (1, 1), (\mathcal{W}_{i}, 1) \\
 (\mathcal{U}_{i}, 1), (\mathcal{K}_{i}, 1), (0, 1) \\
\end{array}
\right] dx
\\
\nonumber
&=\varphi^{-(u+v+w_{z}+1)}
\\
&\times H_{1,0:1,1:2,3}^{1,0:1,1:2,1}\biggl[\begin{array}{c}
J_{11}\\
J_{2}\\
\end{array}\biggl | \begin{array}{c}
(0,1)\\
(0,1)\\
\end{array}\biggl | \begin{array}{c}
(1, 1), (\mathcal{W}_{i}, 1) \\
 (\mathcal{U}_{i}, 1), (\mathcal{K}_{i}, 1), (0, 1)\\
\end{array}\biggl | J_{6},J_{7}\biggl],
\end{align}
where $J_{11}=(-u-v-w_{z};1,\mathcal{V}_{i})$.}
Similar to $\mathcal{A}_{2}$, converting $\frac{1}{1+x}$ into Meijer's $G$ function and performing integration with the help of \cite[Eq.~(2.24.1.1)]{Calculationbook02}, $\mathcal{B}_{4}$ is derived as
\textcolor{black}{\begin{align}
\nonumber
\mathcal{B}_{4}&=\int_{0}^{\infty}\frac{x^{w_{z}}}{1+x}G_{2,3}^{2,1}\left[\mathcal{N}_{i}x^{\mathcal{V}_{i}}\biggl |
\begin{array}{c}
 1, \mathcal{W}_{i} \\
 \mathcal{U}_{i}, \mathcal{K}_{i}, 0 \\
\end{array}
\right]dx
\\
\nonumber
&=\int_{0}^{\infty}x^{w_{z}} G_{1,1}^{1,1}\left[x\biggl |
\begin{array}{c}
0 \\
0 \\
\end{array}
\right]G_{2,3}^{2,1}\left[\mathcal{N}_{i}x^{\mathcal{V}_{i}}\biggl |
\begin{array}{c}
 1, \mathcal{W}_{i} \\
 \mathcal{U}_{i}, \mathcal{K}_{i}, 0 \\
\end{array}
\right] dx
\\
\nonumber
&=\frac{1}{2\pi^{(\mathcal{V}_{i}-1)}}
\\
&\times G_{2+\mathcal{V}_{i},3+\mathcal{V}_{i}}^{2+\mathcal{V}_{i}, 1+\mathcal{V}_{i}}\left[\mathcal{N}_{i}\biggl |
\begin{array}{c}
 1, \Delta(\mathcal{V}_{i},-w_{z}), \Delta(1, \mathcal{W}_{i}) \\
\Delta(1, \mathcal{U}_{i}), \Delta(1, \mathcal{K}_{i}),  \Delta(\mathcal{V}_{i},-w_{z}), 0 \\
\end{array}
\right].
\end{align}}
{\color{black}It is noteworthy that although \eqref{eqn:asc2} is represented here as an infinite series, it converges very rapidly to its accurate results within twenty terms \cite{jameson2016incomplete}.}
\subsection{Secrecy Outage Probability Analysis}
\begin{figure*}[!t]
\textcolor{black}
{\begin{align}
\vspace{-5mm}
\label{eqn:sop3}
\nonumber
SOP_{L}&=\frac{h_{e}^{N_{e}\mu_{e}}}{H_{e}^{N_{e}\mu_{e}}\Gamma(N_{e}\mu_{e})} \sum_{\delta=1}^{2}\sum_{\theta=0}^{N_{e}\mu_{e}-1}D_{\delta,\theta}\biggl[\sum_{m=0}^{N_{S}} \binom{N_{S}}{m}\sum_{n=0}^{m} \binom{m}{n}\sum_{u=0}^{(m-n)(N_{r}\mu_{r}-1)}\sum_{v=0}^{n(N_{r}\mu_{r}-1)}R\sigma^{u+v}
\\
\nonumber
&\times \biggl(\mathcal{B}_{1}-\sum_{i=1}^{2}\mathcal{S}_{i}\mathcal{V}_{i}^{\omega_{B_{1}}-\frac{1}{2}} (\omega_{B_{2}})^{-\omega_{B_{1}}} (2\pi)^{-(\frac{\mathcal{V}_{i}-1}{2})}G_{2+\mathcal{V}_{i},3}^{2,1+\mathcal{V}_{i}}\left[\frac{\mathcal{N}_{i}(\sigma)^{\mathcal{V}_{i}}}{(\omega_{B_{2}})^{\mathcal{V}_{i}}\mathcal{V}_{i}^{-\mathcal{V}_{i}}}\biggl |
\begin{array}{c}
1, \omega_{B_{4}}, \Delta(1, \mathcal{W}_{i}) \\
\Delta(1, \mathcal{U}_{i}), \Delta(1, \mathcal{K}_{i}), 0 \\
\end{array}
\right]\biggl)
\\
& +\sum_{i=1}^{2}\mathcal{S}_{i}\frac{\mathcal{V}_{i}^{N_{e}\mu_{e}-\theta-\frac{1}{2}}}{(2\pi)^{\frac{\mathcal{V}_{i}-1}{2}}}(\Omega_{\delta})^{-(N_{e}\mu_{e}-\theta)} G_{2+\mathcal{V}_{i},3}^{2,1+\mathcal{V}_{i}}\left[\frac{\mathcal{N}_{i}(\sigma)^{\mathcal{V}_{i}}}{(\Omega_{\delta})^{\mathcal{V}_{i}}\mathcal{V}_{i}^{-\mathcal{V}_{i}}}\biggl |
\begin{array}{c}
1, \omega_{B_{3}}, \Delta(1, \mathcal{W}_{i}) \\
\Delta(1, \mathcal{U}_{i}), \Delta(1, \mathcal{K}_{i}), 0 \\
\end{array}
\right]\biggl].
\end{align}}
\hrulefill
\end{figure*}

SOP is another fundamental performance metric that scrutinizes the secrecy behaviour of a system in presence of potential eavesdropper. It can be described as the probability at which instantaneous secrecy capacity ($C_{sc}$) downfalls than predecided target secrecy rate ($\varpi_{o}$)  \cite{wyner1975wire}. So, the SOP is defined mathematically as  \cite[Eq.~(14)]{[49]lei2016}
\begin{align}\label{eqn:sop}
\nonumber
SOP&= \mathrm{Pr}\left\{C_{sc}\leq \varpi_{o}\right\}= \mathrm{Pr}\left\{\gamma_{f}\leq\sigma\gamma_{e}+\sigma-1\right\}
\\
&=\int_{0}^{\infty}F_{\gamma_{f}}(\sigma x+\sigma-1)f_{\gamma_{e}}(x)dx,
\end{align}
where $\sigma=2^{\varpi_{o}}$. Likely to \cite[Eq.~(6)]{[50]lei2015}, we determine the analytical expression of lower-bound SOP as
\begin{align}\label{eqn:sop1}
\nonumber
SOP\geq SOP_{L}&= \mathrm{Pr}\left\{\gamma_{f}\leq\sigma\gamma_{e}\right\}
\\
&=\int_{0}^{\infty}F_{\gamma_{f}}(\sigma x)f_{\gamma_{e}}(x)dx.
\end{align}
By substituting \eqref{eqn:dualhopcdf01} and \eqref{eqn:eavpdf} into \eqref{eqn:sop1}, the expression of SOP in closed-form is developed as \eqref{eqn:sop3}, where the derivation of three distinct integral terms $\mathcal{S}_{1}$, $\mathcal{S}_{2}$, and $\mathcal{S}_{3}$ are calculated as follows.

On performing integration by means of \cite[Eq.(3.326.2)]{jeffrey2007table}, $\mathcal{S}_{1}$ is derived as
\begin{align}
\setcounter{equation}{36}
\mathcal{S}_{1}&=\int_{0}^{\infty}x^{\omega_{B_{1}}-1}e^{-\omega_{B_{2}} x} dx
=\frac{\Gamma(\omega_{B_{1}})}{(\omega_{B_{2}})^{\omega_{B_{1}}}},
\end{align}
where $\omega_{B_{1}}=u+v+N_{e}\mu_{e}-\theta$ and $\omega_{B_{2}}=\Omega_{\delta}+\sigma \varphi$.
Following similar mathematical approach as for $\mathcal{A}_{2}$ and performing integration with the help of identity \cite[Eq.~(2.24.1.1)]{Calculationbook02}, $\mathcal{S}_{2}$ is obtained as
\textcolor{black}{\begin{align}
\nonumber
\mathcal{S}_{2}&=\int_{0}^{\infty}x^{N_{e}\mu_{e}-\theta-1}e^{-\Omega_{\delta} x}
G_{2,3}^{2,1}\left[\mathcal{N}_{i}(x\sigma)^{\mathcal{V}_{i}}\biggl |
\begin{array}{c}
 1, \mathcal{W}_{i} \\
 \mathcal{U}_{i}, \mathcal{K}_{i}, 0 \\
\end{array}
\right]dx
\\
\nonumber
&=\int_{0}^{\infty}x^{N_{e}\mu_{e}-\theta-1}G_{0,1}^{1,0}\left[\Omega_{\delta}x\biggl |
\begin{array}{c}
- \\
0 \\
\end{array}
\right]
\\
\nonumber
& \times G_{2,3}^{2,1}\left[\mathcal{N}_{i}(x\sigma)^{\mathcal{V}_{i}}\biggl |
\begin{array}{c}
 1, \mathcal{W}_{i} \\
 \mathcal{U}_{i}, \mathcal{K}_{i}, 0 \\
\end{array}
\right]dx
\\
\nonumber
&=\frac{\mathcal{V}_{i}^{N_{e}\mu_{e}-\theta-\frac{1}{2}}}{(\Omega_{\delta})^{N_{e}\mu_{e}-\theta}(2\pi)^{\frac{\mathcal{V}_{i}-1}{2}}}
\\
&\times G_{2+\mathcal{V}_{i},3}^{2,1+\mathcal{V}_{i}}\left[\frac{\mathcal{N}_{i}(\sigma)^{\mathcal{V}_{i}}}{(\Omega_{\delta})^{\mathcal{V}_{i}}\mathcal{V}_{i}^{-\mathcal{V}_{i}}}\biggl |
\begin{array}{c}
1, \omega_{B_{3}}, \Delta(1, \mathcal{W}_{i}) \\
\Delta(1, \mathcal{U}_{i}), \Delta(1, \mathcal{K}_{i}), 0 \\
\end{array}
\right],
\end{align}
where $\omega_{B_{3}}=\Delta(\mathcal{V}_{i},1-N_{e}\mu_{e}+\theta)$.
Based on $\mathcal{S}_{2}$, $\mathcal{S}_{3}$ is obtained as \begin{align}
\nonumber
\mathcal{S}_{3}&=\int_{0}^{\infty}x^{\omega_{B_{1}}-1}e^{-\omega_{B_{2}} x}
G_{2,3}^{2,1}\left[\mathcal{N}_{i}(x\sigma)^{\mathcal{V}_{i}}\biggl |
\begin{array}{c}
 1, \mathcal{W}_{i} \\
 \mathcal{U}_{i}, \mathcal{K}_{i}, 0 \\
\end{array}
\right]dx
\\
\nonumber
&=\int_{0}^{\infty}x^{\omega_{B_{1}}-1}G_{0,1}^{1,0}\left[\omega_{B_{2}}x\biggl |
\begin{array}{c}
- \\
0 \\
\end{array}
\right]
\\
\nonumber
&\times G_{2,3}^{2,1}\left[\mathcal{N}_{i}(x\sigma)^{\mathcal{V}_{i}}\biggl |
\begin{array}{c}
 1, \mathcal{W}_{i} \\
 \mathcal{U}_{i}, \mathcal{K}_{i}, 0 \\
\end{array}
\right]dx
\\
\nonumber
&=\frac{\mathcal{V}_{i}^{\omega_{B_{1}}-\frac{1}{2}}}{(\omega_{B_{2}})^{\omega_{B_{1}}}(2\pi)^{\frac{\mathcal{V}_{i}-1}{2}}}
\\
&\times G_{2+\mathcal{V}_{i},3}^{2,1+\mathcal{V}_{i}}\left[\frac{\mathcal{N}_{i}(\sigma)^{\mathcal{V}_{i}}}{(\omega_{B_{2}})^{\mathcal{V}_{i}}\mathcal{V}_{i}^{-\mathcal{V}_{i}}}\biggl |
\begin{array}{c}
1, \omega_{B_{4}}, \Delta(1, \mathcal{W}_{i}) \\
\Delta(1, \mathcal{U}_{i}), \Delta(1, \mathcal{K}_{i}), 0 \\
\end{array}
\right],
\end{align}
where $\omega_{B_{4}}=\Delta(\mathcal{V}_{i},1-\omega_{B_{1}})$.}


\subsection{Probability of Strictly Positive Secrecy Capacity Analysis}

In the wiretapped model, probability of SPSC is an essential factor guaranteeing uninterrupted communication that is only possible if confidentiality capability remains positive ($\gamma_{f}>\gamma_{e}$). It is important to note although the closed-form expression of SPSC can be easily determined from SOP expression, it has a different physical meaning that inspects the existence of $C_{sc}$. Thus, probability of SPSC can be mathematically expressed as  \cite[Eq.~(25)]{islam2020secrecy}
\begin{align}
\nonumber
SPSC&=\Pr(C_{sc}>0)=1-\Pr\left\{C_{sc}\leq0\right\}
\\
\label{a23}
&=1-SOP_{L}\left(\varpi_{o}=0\right).
\end{align}
With the help of \eqref{a23}, the probability of SPSC can be obtained. Therefore, substituting $\varpi_{o}=0$ in \eqref{eqn:sop3}, the analytical expression of probability of SPSC is finally determined as expressed in \eqref{eqn:spscfinal}.
\begin{figure*}[!t]
\textcolor{black}{\begin{align}
\vspace{-5mm}
\label{eqn:spscfinal}
\nonumber
SPSC&=1-\frac{h_{e}^{N_{e}\mu_{e}}}{H_{e}^{N_{e}\mu_{e}}\Gamma(N_{e}\mu_{e})} \sum_{\delta=1}^{2}\sum_{\theta=0}^{N_{e}\mu_{e}-1}D_{\delta,\theta}\biggl[\sum_{m=0}^{N_{S}} \binom{N_{S}}{m}\sum_{n=0}^{m} \binom{m}{n}\sum_{u=0}^{(m-n)(N_{r}\mu_{r}-1)}\sum_{v=0}^{n(N_{r}\mu_{r}-1)}R\biggl(\Gamma(\omega_{B_{1}})(\omega_{B_{2}})^{-\omega_{B_{1}}}
\\
\nonumber
&-\sum_{i=1}^{2}\mathcal{S}_{i}\frac{\mathcal{V}_{i}^{\omega_{B_{1}}-\frac{1}{2}}}{(\omega_{B_{2}})^{\omega_{B_{1}}}}
(2\pi)^{-(\frac{\mathcal{V}_{i}-1}{2})}G_{2+\mathcal{V}_{i},3}^{2,1+\mathcal{V}_{i}}\left[\frac{\mathcal{N}_{i}}{(\omega_{B_{2}})^{\mathcal{V}_{i}}\mathcal{V}_{i}^{-\mathcal{V}_{i}}}\biggl |
\begin{array}{c}
1, \omega_{B_{4}}, \Delta(1, \mathcal{W}_{i}) \\
\Delta(1, \mathcal{U}_{i}), \Delta(1, \mathcal{K}_{i}), 0 \\
\end{array}
\right]\biggl)-\sum_{i=1}^{2}\mathcal{S}_{i}\frac{\mathcal{V}_{i}^{N_{e}\mu_{e}-\theta-\frac{1}{2}}}{(\Omega_{\delta})^{N_{e}\mu_{e}-\theta}(2\pi)^{\frac{\mathcal{V}_{i}-1}{2}}}
\\
&\times
G_{2+\mathcal{V}_{i},3}^{2,1+\mathcal{V}_{i}}\left[\frac{\mathcal{N}_{i}}{(\Omega_{\delta})^{\mathcal{V}_{i}}\mathcal{V}_{i}^{-\mathcal{V}_{i}}}\biggl |
\begin{array}{c}
1, \omega_{B_{3}}, \Delta(1, \mathcal{W}_{i}) \\
\Delta(1, \mathcal{U}_{i}), \Delta(1, \mathcal{K}_{i}), 0 \\
\end{array}
\right]\biggl].
\end{align}}
\hrulefill
\end{figure*}


\section{Numerical Results}

For a mixed dual-hop system configuration, performance metrics, i.e., ASC, SOP, and probability of SPSC are analyzed graphically in this section to investigate the influence of system parameters on the secrecy performance of the proposed framework. In order to verify the accuracy of developed novel expressions in \eqref{eqn:asc1}, \eqref{eqn:asc2}, \eqref{eqn:sop3}, and \eqref{eqn:spscfinal}, MC simulations are performed by generating $10^{6}$ independent random samples in MATLAB and we notice a close match between analytical and simulation results. Moreover, bivariate Fox's $H$ function can not directly be evaluated using Mathematica or Maple since there is no built-in function for that purpose and hence \cite[Table~I]{lei2017secrecy} is utilized to implement this function in Mathematica. In addition, Format $1$ is considered for the first hop while various UWT conditions (i.e. weak to severe) that are applicable to both fresh and salty waters under temperature gradients and thermally uniform conditions are taken from \cite[Table~I,II]{zedini2019unified} applicable to the second hop.
\begin{figure}[!h]
\vspace{-10mm}
    \centerline{\includegraphics[width=0.65\textwidth,angle =0]{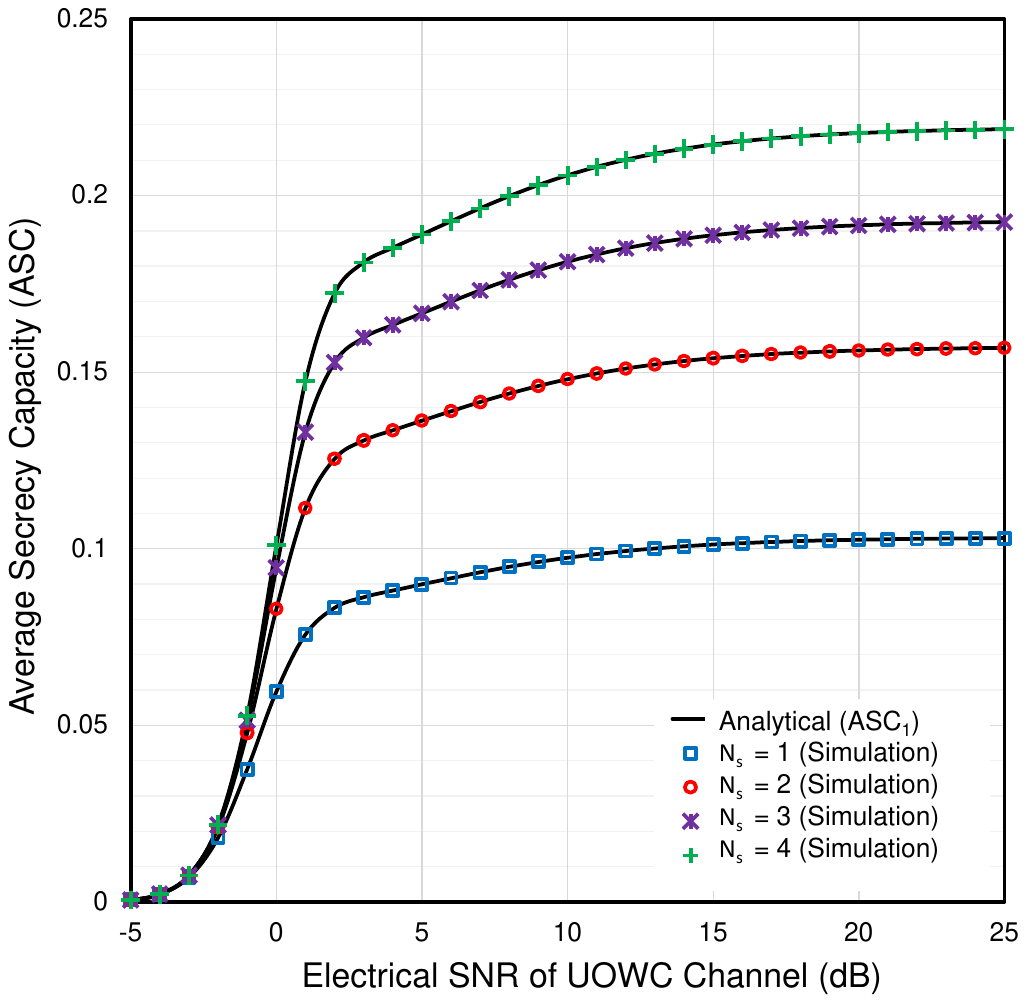}}
        \vspace{-10mm }
    \caption{
         The $ASC_{1}$ versus $\Psi_{r}$ for selected values of $N_{s}$ with $N_{r}=N_{e}=2$, $\eta_{r}=\eta_{e}=2.2$, $\mu_{r}=\mu_{e}=2$, $h = 2.4$, $l = 0.05$, $r=1$, and $\varphi_{r}=\varphi_{e}=0$ dB in temperature gradient water.
    }
    \label{fig02}
\end{figure}

\begin{figure}[!h]
\vspace{-10mm}
    \centerline{\includegraphics[width=0.65\textwidth,angle =0]{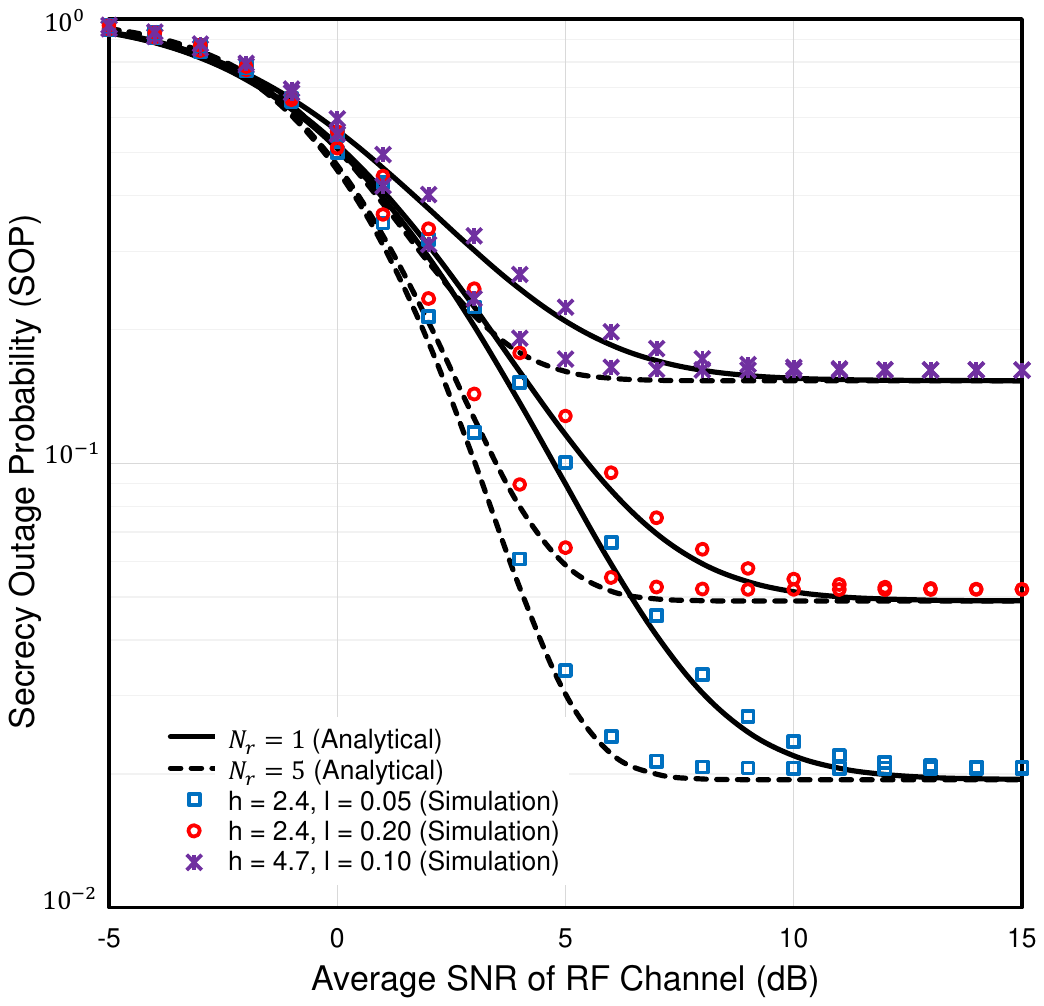}}
        \vspace{-10mm }
    \caption{
         The SOP versus $\varphi_{r}$ for selected values of $h$, $l$, and $N_{r}$ with $N_{s}=N_{e}=1$, $\eta_{r}=\eta_{e}=2.2$, $\mu_{r}=\mu_{e}=2$, $r=1$, $\varphi_{e}=0$ dB, $\varphi_{d}=15$ dB, and $\varpi_{o}=0.01$ bits/s/Hz in temperature gradient water.
         }
    \label{fig03}
\end{figure}

\begin{figure}[!h]
\vspace{-30mm}
    \centerline{\includegraphics[width=0.45\textwidth,angle =0]{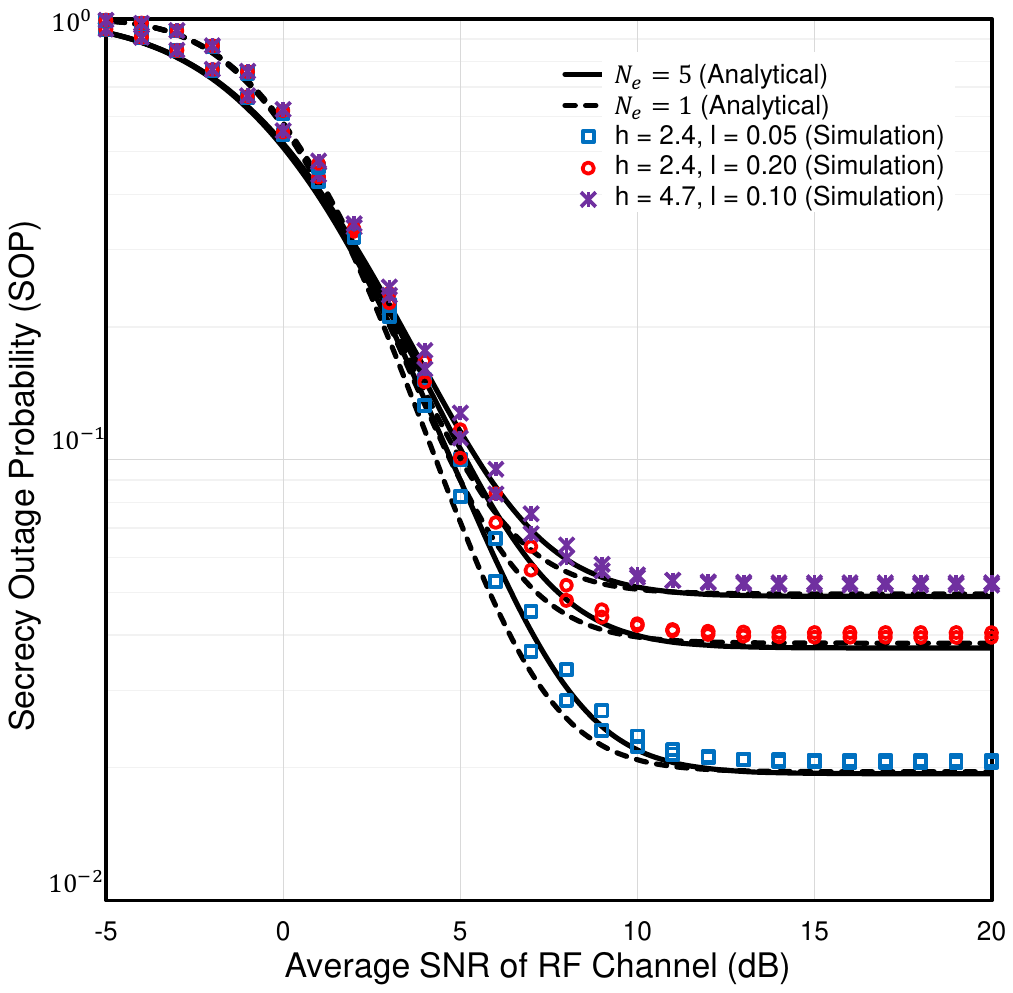}}
        \vspace{-25mm }
    \caption{The SOP versus $\varphi_{r}$ for selected values of $h$, $l$, and $N_{e}$ with $N_{s}=N_{r}=1$, $\eta_{r}=\eta_{e}=2.2$, $\mu_{r}=\mu_{e}=2$, $r=1$, $\varphi_{e}=0$ dB, $\varphi_{d}=15$ dB, and $\varpi_{o}=0.01$ bits/s/Hz in temperature gradient water.}
    \label{fig04}
\end{figure}

In Fig. \ref{fig02}, ASC is plotted as a function of $\Psi_{r}$ for different numbers of $N_{s}$. Here, the main goal is to inspect the influence of transmit antennas on secrecy performance for selected values of $N_{s}$. It can be observed clearly that ASC proportionally increases with an increase of $N_{s}$. It is as expected because an increase in $N_{s}$ increases the probability of obtaining the strongest transmit signal (i.e. signal with highest SNR among $N_{s}$ antennas) from source as testified in \cite{yang2018physical}.

\begin{figure}[!h]
\vspace{-10mm}
    \centerline{\includegraphics[width=0.65\textwidth,angle =0]{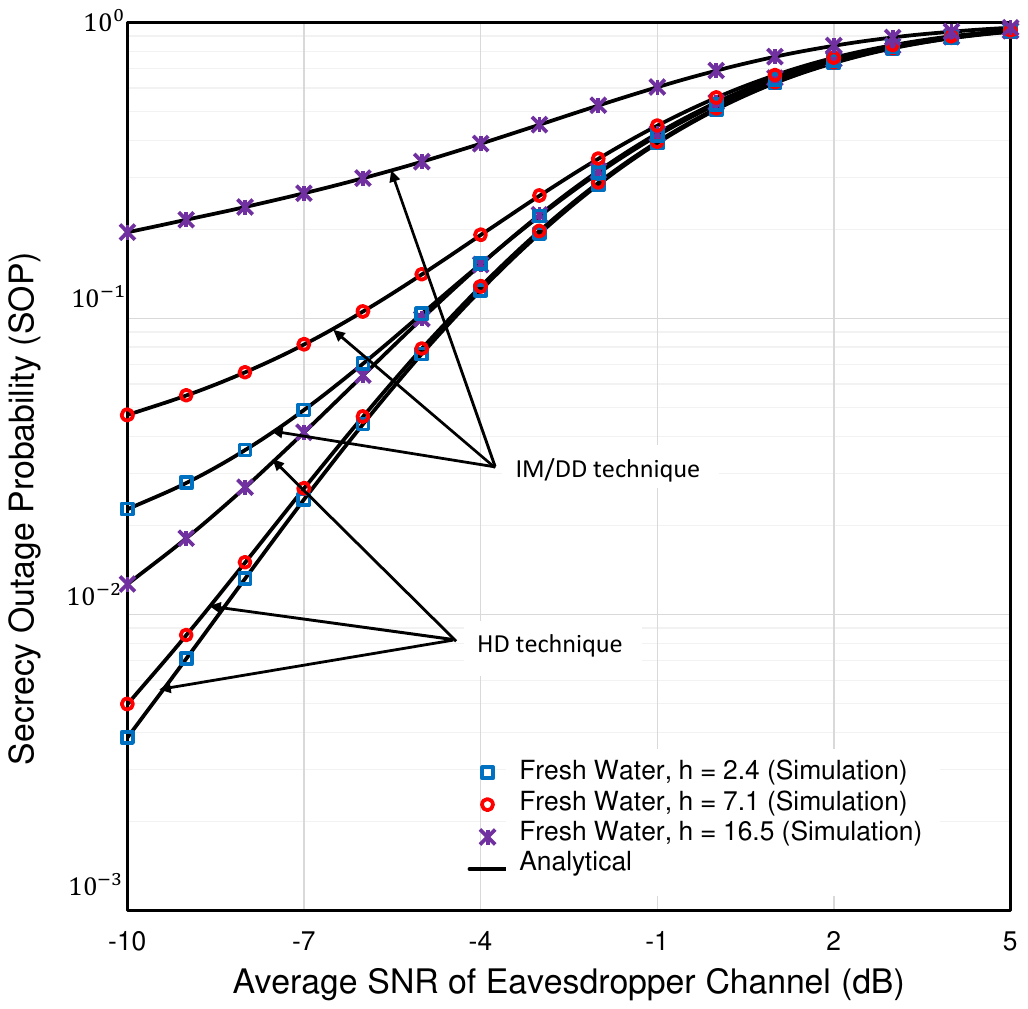}}
        \vspace{-10mm }
    \caption{
         The SOP versus $\varphi_{e}$ for selected values of $h$, $l$, and $r$ with $N_{s}= N_{r}=N_{e}=1$, $\eta_{r}=\eta_{e}=2.2$, $\mu_{r}=\mu_{e}=2$, $\varphi_{r}=0$ dB, $\varphi_{d}=15$ dB, and $\varpi_{o}=0.01$ bits/s/Hz in thermally uniform fresh water.
    }
    \label{fig05}
\end{figure}

\begin{figure}[!h]
\vspace{-10mm}
    \centerline{\includegraphics[width=0.65\textwidth,angle =0]{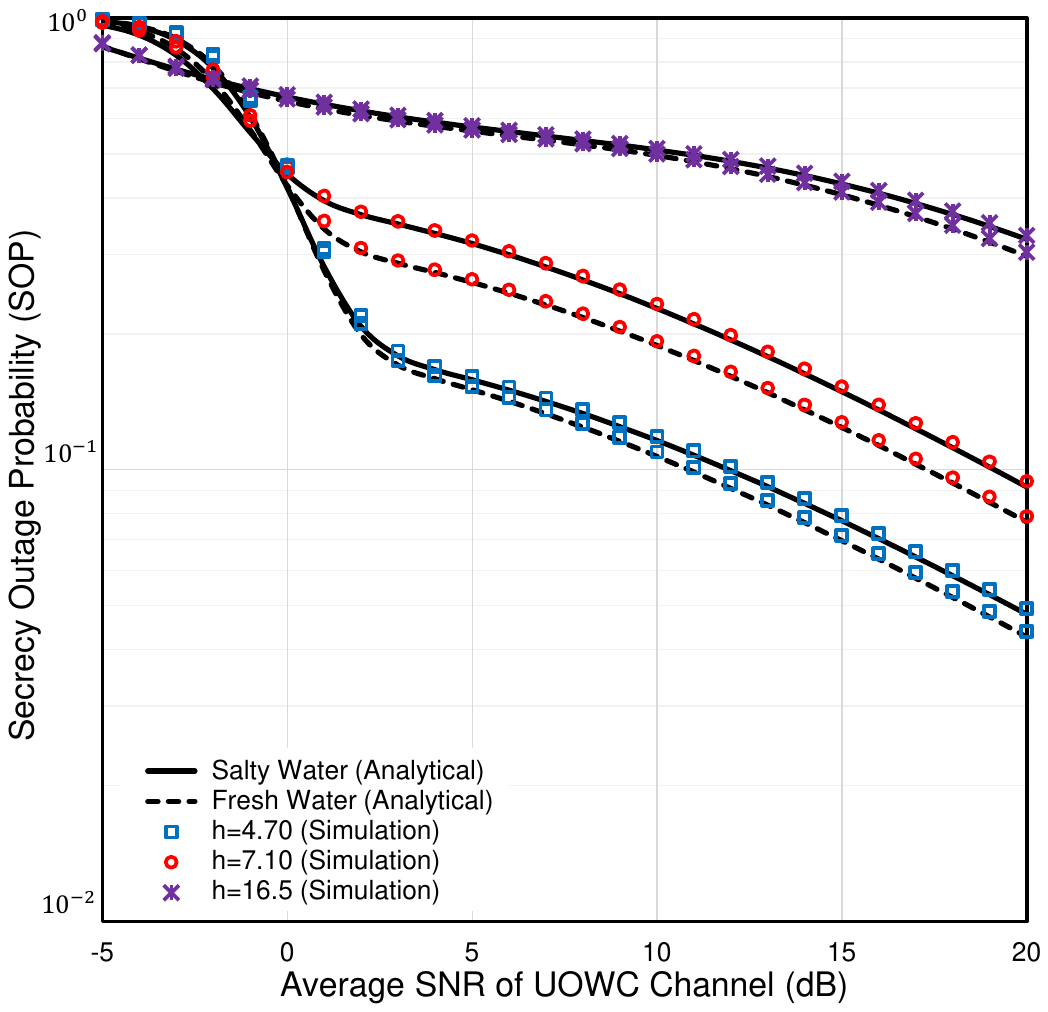}}
        \vspace{-10mm }
    \caption{
         The SOP versus $\varphi_{d}$ for selected values of $h$ with $N_{s}=N_{r}=N_{e}=2$, $\eta_{r}=\eta_{e}=2.2$, $\mu_{r}=\mu_{e}=2$, $r=2$, $\varphi_{r}=20$ dB, $\varphi_{e}=0$ dB, and $\varpi_{o}=0.01$ bits/s/Hz in thermally uniform water.}
    \label{fig06}
\end{figure}

The impact of $N_{r}$ and $N_{e}$ under various turbulence conditions for thermally uniform scenarios are demonstrated in Figs. \ref{fig03} and \ref{fig04}, respectively, where SOP is plotted against $\varphi_{r}$ for a particular number of $N_{s}$. Fig. \ref{fig03} demonstrates that larger the value of $N_{r}$, lower the SOP is. It happens because the MRC diversity at $R$ improves the $S-R$ link significantly, which helps to attain a better SOP performance. On the other hand, Fig. \ref{fig04} reveals that SOP performance with a lower $N_{e}$ is superior to that of a higher one. This is expected due to the fact that increasing $N_{e}$ strengthens the $S-E$ link by enhancing receive diversity at $E$. Notably, an SOP floor is visible in the figures because the system's secrecy performance is always dominated by the RF link (first hop). Moreover, it is important to mention that as the UWT scenario gets severe (i.e. weaker to strong conditions with an increase of air bubbles level and temperature gradients in UOWC link), secrecy characteristics get degraded significantly. Similar results are demonstrated in \cite{zedini2019unified} that validates our analysis.

Fig. \ref{fig05} presents SOP versus $\varphi_{e}$ under various turbulence scenarios in the case of a thermally uniform and freshwater environment.
\begin{figure}[!h]
\vspace{-30mm}
    \centerline{\includegraphics[width=0.45\textwidth,angle =0]{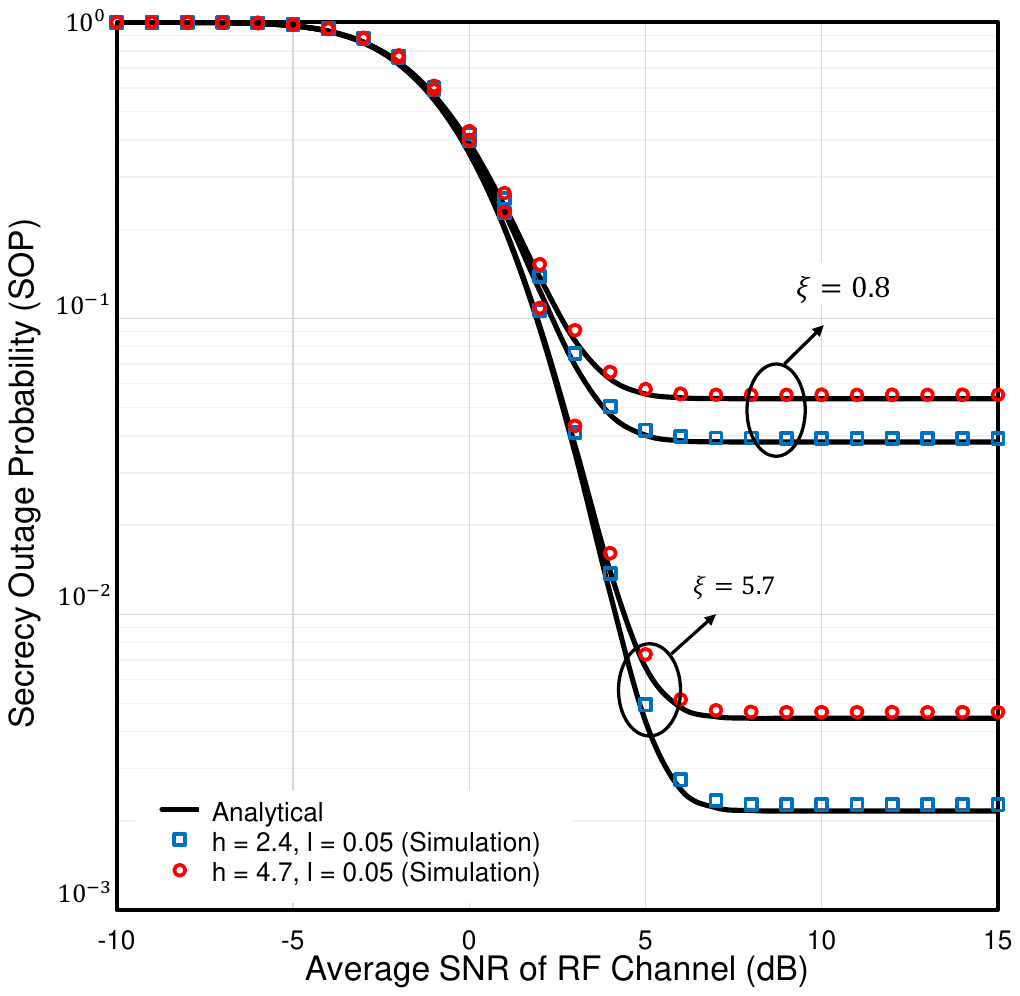}}
        \vspace{-25mm }
    \caption{
         {\color{black}The SOP versus $\varphi_{r}$ for selected values of $h$, $l$, and $\xi$ with $N_{s}=N_{r}=N_{e}=2$, $\eta_{r}=\eta_{e}=2.2$, $\mu_{r}=\mu_{e}=2$, $r=1$, $\varphi_{e}=0$ dB, $\varphi_{d}=25$ dB, and $\varpi_{o}=0.01$ bits/s/Hz in temperature gradient water.}}
    \label{fig10}
\end{figure}

\begin{figure}[ht]
\vspace{-4mm}
    \centerline{\includegraphics[width=0.38\textwidth,angle =270]{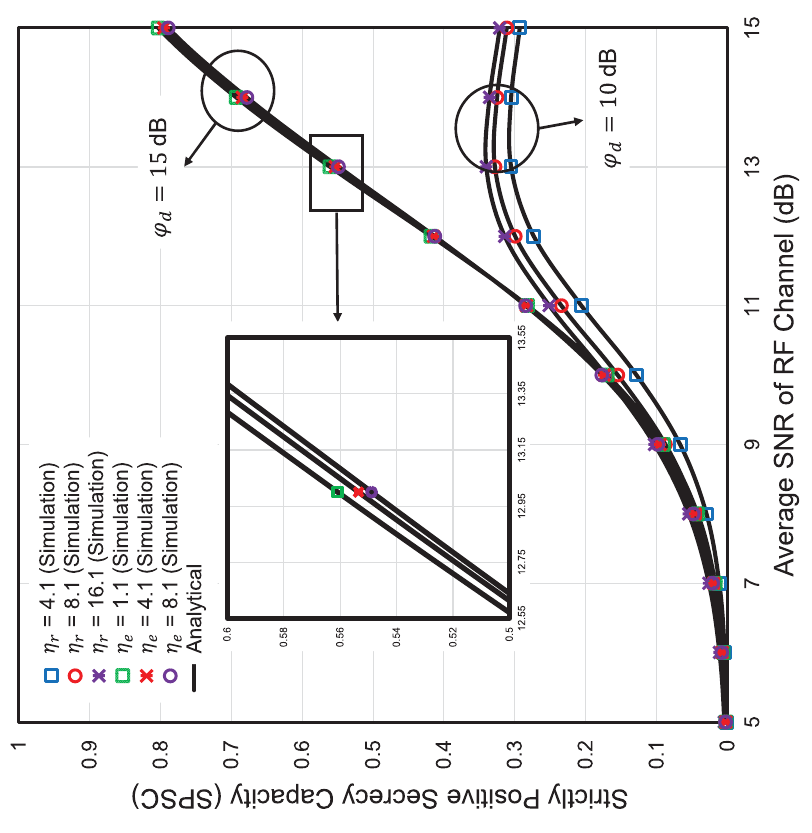}}
        \vspace{0mm }
    \caption{
         The SPSC versus $\varphi_{r}$ for selected values of $\eta_{r}$ and $\eta_{e}$ with $N_{s}=N_{r}=N_{e}=2$, $\mu_{r}=\mu_{e}=2$, $h = 2.4$, $l = 0.20$, $r=2$, and $\varphi_{e}=12$ dB in temperature gradient waters.}
    \label{fig07}
\end{figure}
It is observed that SOP increases with an increase in UWT. It is because with an increase of air bubble levels, the received signal experiences strong fluctuations that lead to an increased scintillation index and a degraded SOP performance. In addition, as expected, the HD technique enhances the secrecy performance of the UOWC link more significantly relative to IM/DD technique that clearly indicates that the HD technique has a better capability to overcome UWT than IM/DD technique. Authors in \cite{lei2020performance} also demonstrated some numerical outcomes regarding detection techniques similar to the proposed model that strengthen this analysis. Moreover, close agreement between analytical and simulation results proves the accuracy and validity of the derived closed-form expressions.
\begin{figure}[!h]
\vspace{-30mm}
    \centerline{\includegraphics[width=0.45\textwidth,angle =0]{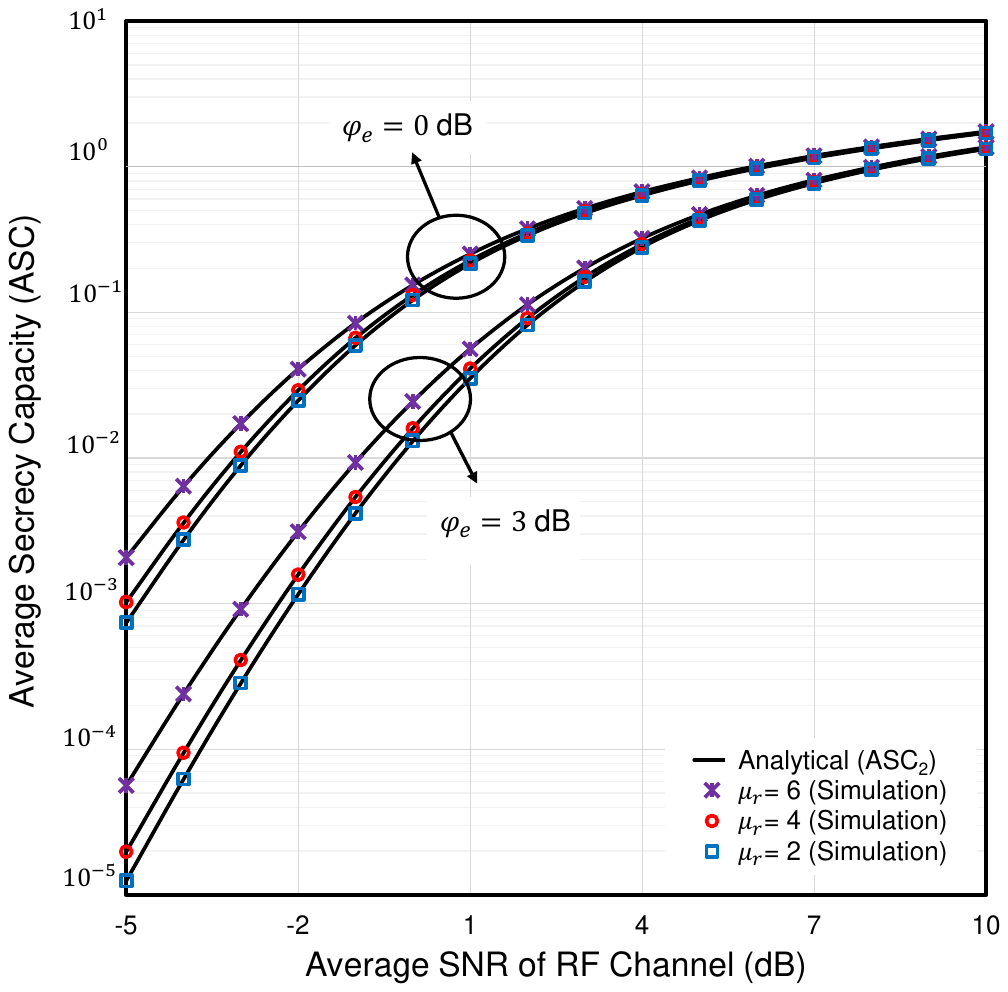}}
        \vspace{-20mm }
    \caption{
         The $ASC_{2}$ versus $\varphi_{r}$  for selected values of $\mu_{r}$ with $N_{s}=N_{r}=N_{e}=2$, $\eta_{r}=\eta_{e}=2.2$, $\mu_{e}=2$, $h=2.4$, $l= 0.05$, $\xi=5.7$, $r=1$, and $\varphi_{d}=15$ dB in temperature gradient water.
    }
    \label{fig08}
\end{figure}
Besides air bubble levels and temperature gradients, salinity also induces turbulence in the UOWC link. This phenomenon is demonstrated in Fig. \ref{fig06} by depicting SOP as a function of $\varphi_{d}$ with various levels of air bubbles under uniform temperature.

We can observe that fresh waters demonstrate better SOP performance relative to salty water. But the impact of water salinity is less severe than the impacts of air bubbles. Similar outcomes were also reported in \cite{zedini2019unified}.
\begin{figure}[!h]
\vspace{-12mm}
    \centerline{\includegraphics[width=0.60\textwidth,angle =0]{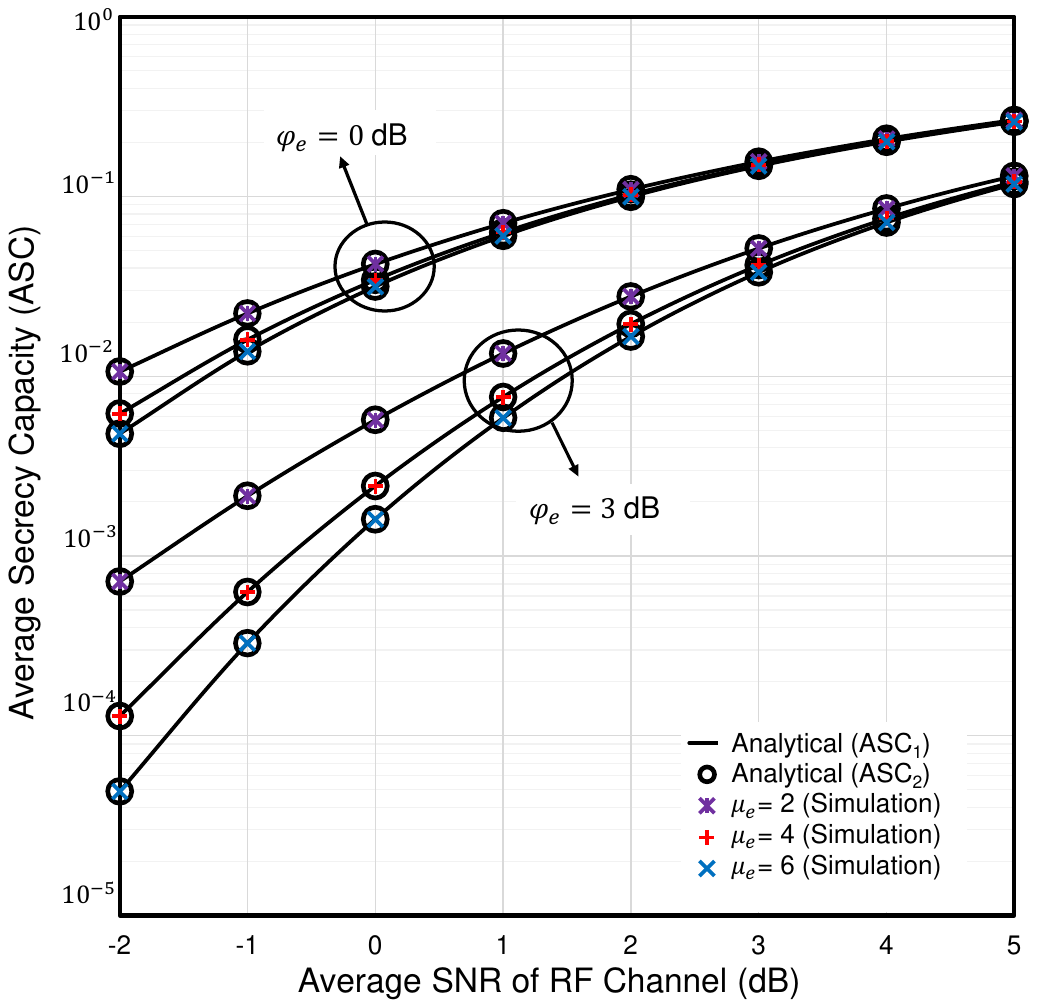}}
        \vspace{-08mm }
    \caption{
         The $ASC_{1}$ \& $ASC_{2}$ versus $\varphi_{r}$ for selected values of $\mu_{e}$ with $N_{s}=1$, $N_{r}=N_{e}=2$, $\eta_{r}=\eta_{e}=2.2$, $\mu_{r}=2$, $h = 16.5$, $l = 0.22$, $r=2$, and $\varphi_{d}=15$ dB in temperature gradient water.}
    \label{fig09}
\end{figure}

{\color{black}In Fig. \ref{fig10}, the SOP is demonstrated as a function of $\varphi_{r}$ with a view of observing impacts of pointing error under temperature gradient water. It can be noted that higher the pointing error (i.e. lower $\xi$), lower the secrecy performance (i.e. higher SOP). This is because higher $\xi$ signifies pronounced pointing accuracy \cite{li2021performance}.}


Figure \ref{fig07} presents SPSC versus $\varphi_{r}$ while Figs. \ref{fig08}-\ref{fig09} demonstrates ASC versus $\varphi_{r}$ to quantify influences of RF fading parameters.
\begin{figure}[!ht]
\vspace{-30mm}
    \centerline{\includegraphics[width=0.45\textwidth,angle =0]{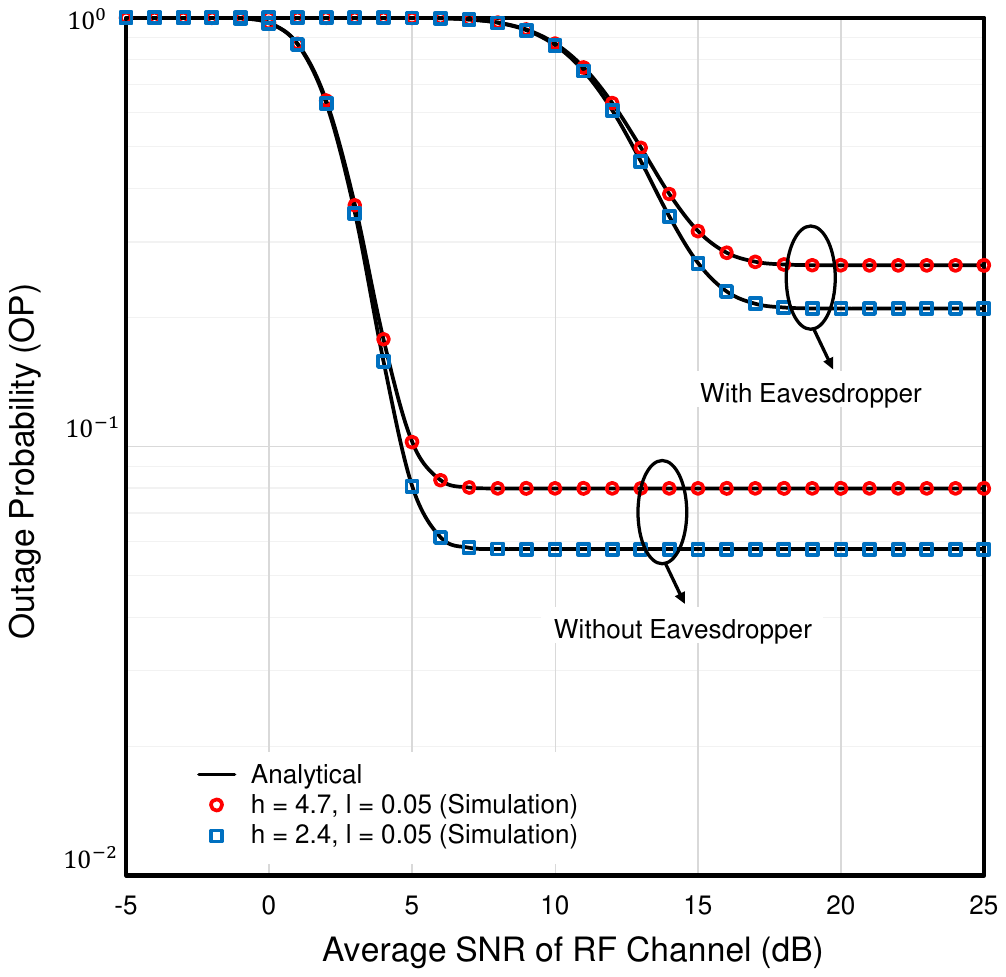}}
        \vspace{-25mm }
    \caption{{\color{black}
         The OP versus $\varphi_{r}$ for selected values of $h$, $l$, and $\xi$ with $N_{s}=N_{r}=N_{e}=2$, $\eta_{r}=\eta_{e}=2.2$, $\mu_{r}=\mu_{e}=2$, $r=1$, $\varphi_{e}=10$ dB, $\varphi_{d}=25$ dB, and $\varpi_{o}=0.01$ bits/s/Hz in temperature gradient water.}}
    \label{fig11}
\end{figure}

It is observed that due to the increase of $\eta_{r}$ and $\mu_{r}$, a better secrecy behavior of the proposed scenario is exhibited. This performance is desired because the fading severity of the $S-R$ link decreases with $\eta_{r}$ and $\mu_{r}$. On the other hand, the higher the value of $\eta_{e}$ and $\mu_{e}$, the larger is the possibility of eavesdropping since in such a case the wiretap link will be strengthened. The same sort of results is also observed in \cite{yang2018physical}. It can also be noted that both expressions of ASC i.e. ASC$_{1}$ and ASC$_{2}$ yields the same results.  

\begin{figure}[!ht]
\vspace{-10mm}
    \centerline{\includegraphics[width=0.60\textwidth,angle =0]{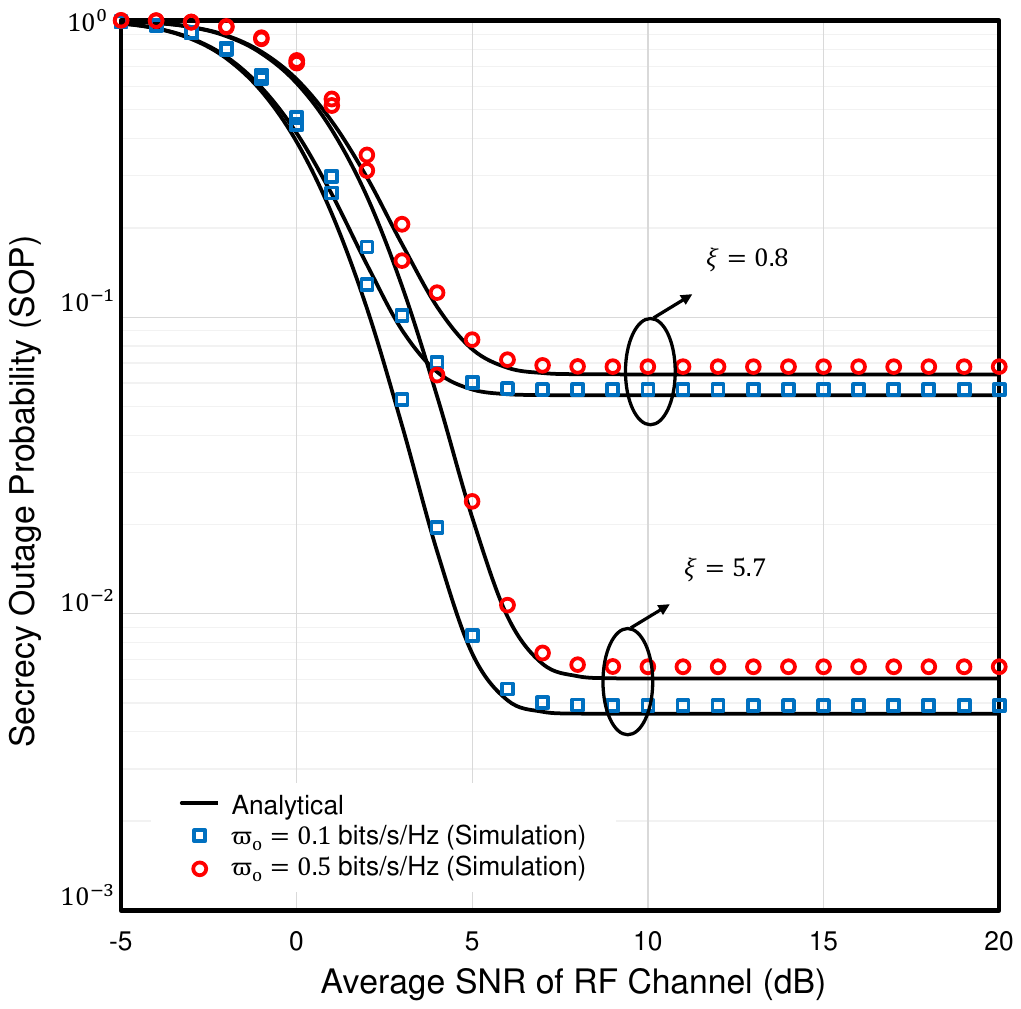}}
        \vspace{-10mm}
    \caption{{\color{black}
         The SOP versus $\varphi_{r}$ for selected values of $\varpi_{o}$ with $N_{s}=N_{r}=N_{e}=2$, $\eta_{r}=\eta_{e}=2.2$, $\mu_{r}=\mu_{e}=2$, $h=4.7$, $l=0.05$, $r=1$, $\varphi_{e}=0$ dB, and $\varphi_{d}=25$ dB in temperature gradient water.}}
    \label{fig12}
\end{figure}
\textcolor{black}{Figure \ref{fig11} describes the difference between system performances of two cases (i.e. with and without eavesdropping scenarios). It is evident that in absence of any potential eavesdroppers, we need to transmit at a rate greater than $\gamma_{th}$. But PLS technique assumes a secrecy rate $C_{sc}$ (i.e. the rate at which the eavesdroppers are unable to decode the transmitted data) that is defined as the difference between capacity of main and eavesdropper channels and always less than main channel capacity. Hence, in the passive eavesdropping scenario, for same prefixed SNR threshold (i.e. $\gamma_{th}=\varpi_{o}$), outage performance deteriorates relative to that of without eavesdropping case.}

{\color{black} In Figure \ref{fig12}, the impact of target secrecy rate on secure outage performance is demonstrated wherein SOP is plotted against $\varphi_{r}$. It is noted that increase in $\varpi_{o}$ increases the probability of $C_{sc}$ dropping below $\varpi_{o}$ leading to a degraded secrecy performance, as demonstrated in \cite{islam2020secrecy}.}

\section{Conclusion}
In this work, we have investigated the secrecy behavior of a variable gain AF relying based RF-UOWC mixed network where fading, TAS/MRC diversity scheme, and various UWT conditions due to the salty and fresh waters under temperature gradients and thermally uniform underwater environments are taken into consideration. However, assuming all RF and UOWC links are subjected to $\eta$-$\mu$ and mEGG models, respectively, analytical expressions of ASC, SOP, and probability of SPSC are derived in closed-form and then validated via MC simulation. Numerical results reveal that although UWT deteriorates the security performance of the proposed model, it can be significantly enhanced by exploiting TAS/MRC scheme at the main RF link. It is also noteworthy that the HD technique can overcome UWT with better security compared to IM/DD technique.   

\bibliographystyle{IEEEtran}
\bibliography{IEEEabrv,main.bib}

\end{document}